\documentclass[fleqn,usenatbib]{mnras}

\usepackage{newtxtext,newtxmath}
\usepackage[T1]{fontenc}
\DeclareRobustCommand{\VAN}[3]{#2}
\let\VANthebibliography\thebibliography
\def\thebibliography{\DeclareRobustCommand{\VAN}[3]{##3}\VANthebibliography}

\usepackage{graphicx}	% Including figure files
\usepackage{amsmath}	% Advanced maths commands

\usepackage{natbib}
\usepackage{lscape}
\usepackage{longtable}

\newcommand{\feh}{\rm [Fe/H]}
\newcommand{\rgc}{$R_{\rm GC}$}
\newcommand{\rguide}{$R_{\rm guide}$}
\newcommand{\rbirth}{$R_{\rm birth}$}
\newcommand{\dexkpc}{dex\,kpc$^{-1}$}
\newcommand{\kpcgyr}{kpc\,Gyr$^{-1}$}

\title[The Galactic metallicity gradient]{The Galactic metallicity gradient shown by open clusters in the light of radial migration}

\author[M. Netopil et al.]{
Martin Netopil,$^{1}$\thanks{E-mail: mn.netopil@gmail.com}
{\.I}nci Akkaya Oralhan,$^{2}$
 Hikmet \c{C}akmak,$^{3}$
Ra\'ul Michel,$^{4}$
Y\"uksel Karata\c{s}$^{3}$
\\
$^{1}$Kuffner Observatory, Johann-Staud-Stra{\ss}e 10, A-1160 Wien, Austria\\
$^{2}$Department of Astronomy and Space Sciences, Science Faculty, Erciyes University, TR-38039, Kayseri, Turkey\\
$^{3}$Department of Astronomy and Space Sciences, Science Faculty, Istanbul University, TR-34116, \"Universite-Istanbul, Turkey\\
$^{4}$Observatorio Astron\'omico Nacional, Universidad Nacional Aut\'onoma de M\'exico, Apartado Postal 877, C.P. 22800, Ensenada, B.C., M\'exico
}

\date{Accepted 2021 October 10. Received 2021 October 10; in original form 2021 April 20}

\pubyear{2021}

\begin{document}
\label{firstpage}
\pagerange{\pageref{firstpage}--\pageref{lastpage}}
\maketitle

\begin{abstract}
During the last years and decades several individual studies and large-scale spectroscopic surveys significantly improved our knowledge of the Galactic metallicity distribution based on open clusters. The availability of \textit{Gaia} data provided a further step forward in our knowledge. However, still some open issues remain, for example the influence of radial migration on the interpretation of the observed gradients. We used spectroscopic metallicities from individual studies and from the APOGEE survey to compile a sample of 136 open clusters, with a membership verification based on \textit{Gaia} DR2. Additionally, we present photometric metallicity estimates of 14 open clusters in a somewhat outer Galactic region. 
Eight age groups allow us to study the evolution of the metallicity gradient in detail, showing within the errors an almost constant gradient of about $-$0.06 \dexkpc. Furthermore, using the derived gradients and an analysis of the individual objects, we estimate a mean migration rate of 1\,\kpcgyr\ for objects up to about 2\,Gyr. Here, the change of the guiding radius is clearly the main contributor. For older and dynamically hotter objects up to 6\,Gyr we infer a lower migration rate of up to 0.5\,\kpcgyr. The influence of epicyclic excursions increases with age and contributes already about 1\,kpc to the total migration distance after 6\,Gyr. A comparison of our results with available models shows good agreement. However, there is still a lack of a suitable coverage of older objects, future studies are still needed to provide a better sampling in this respect.
 
\end{abstract}

\begin{keywords}
techniques: photometric -- techniques: spectroscopic -- Galaxy: abundances -- open clusters and associations: general
\end{keywords}

%%%%%%%%%%%%%%%%%%%%%%%%%%%%%%%%%%%%%%%%%%%%%%%%%%

%%%%%%%%%%%%%%%%% BODY OF PAPER %%%%%%%%%%%%%%%%%%

\section{Introduction}

Open clusters provide the unique possibility to derive reliable parameters based on their numerous member stars that generally share the same properties.
Open clusters therefore represent invaluable objects for numerous topics related to evolutionary studies. In case of the Galactic metallicity gradient and its evolution, to our best knowledge the first study dates back to \citet{1979ApJS...39..135J}. He used photometric data to identify a metallicity gradient of about $-$0.05 \dexkpc\ and that a younger population shows a shallower gradient than an older one. However, he notes that the age dependence could be also a bias because of the objects' distribution. The existence of a metallicity gradient was consolidated in the last decades even by using larger data sets and by switching to spectroscopic measurements, the evolution of the gradient, however, is still debatable. An overview of literature results for different tracers is presented by \citet{2017A&A...600A..70A}.

Though, still some other open questions remain. For example, is there some evidence of a transition zone in the Milky Way that splits the disc into two regions with different metallicity gradients? \citet{2011MNRAS.417..698L} even conclude two shallow plateaus with a step-like discontinuity. Recently, \citet{2021FrASS...8...62M} identified a flattening of the gradient at a galactocentric distance (\rgc) of about 10\,kpc. Furthermore, they note that the corotation radius, the strongest Galactic resonance close to the solar circle, divides the Galaxy into two parts with independent chemical evolution and causes a gap of the cluster distribution in the range 8.5\,kpc $\leq$ \rgc\ $\leq$ 9.5\,kpc.

The increasing number of available objects allowed to split the available open cluster samples into several age groups to investigate age--metallicity relations in more detail. A shift of the metallicity level with age in young and intermediate-age open clusters is noted in several works \citep[e.g.][]{2016A&A...591A..37J,2016A&A...585A.150N,2020AJ....159..199D,2021MNRAS.503.3279S} or can be inferred from the presented data. It was interpreted e.g. by \citet{2016A&A...585A.150N} as a result of radial mixing. This metallicity shift is not supported by available chemodynamical models \citep[e.g.][]{2013A&A...558A...9M}, but \citet{2016A&A...591A..37J} outlined the variations of different models and their strong sensitivity to the adopted input parameters. \citet{2017A&A...600A..70A} on the other hand proposed a mechanism including radial mixing and cluster disruption to explain the observations of open clusters. They conclude that non- or inward-migrating clusters might be more prone to disruption, leading to an appearance of metal-rich clusters in the solar neighbourhood (which migrated from inwards) and resulting in a steeper gradient for intermediate-age clusters.  However, we also have to deal with some limitation of the chemical abundances itself - the lower metallicity of young objects was recently explained by problems in the spectroscopic analysis \citep[e.g.][]{2020A&A...634A..34B,2021MNRAS.503.3279S}. 

The interpretation of the radial metallicity distribution is generally influenced by several factors. From the observational side these are clearly the reliability of the objects' distance, age and metallicity. However, meanwhile we know that stars (and open clusters) migrate in the Galaxy by a combination of orbital heating and diffusion of orbital angular momentum. These two effects, often also denoted as ``blurring'' for the epicyclic excursions and ``churning'' for changes of the guiding radius \citep[see e.g.][]{2002MNRAS.336..785S,2009MNRAS.396..203S}, result in an additional scatter of the radial metallicity distribution. Its interpretation therefore requires on the one hand accurate cluster parameters, and on the other hand the use of mono-age populations with a sufficiently small age range to obtain some idea about possible variations of the radial migration efficiency by a comparison of these populations.  

\citet{2016A&A...585A.150N} presented a set of mean open cluster metallicities by combining individual spectroscopic measurements on the star level. The sample was divided into two quality categories according to spectroscopic resolution and signal-to-noise ratio, but also includes a set of photometric determinations. The spectroscopic data of 100 objects were used to study the radial metallicity distribution and the age--metallicity relation, and provide a possible observational evidence for radial migration. A main improvement to the work by \citet{2016A&A...585A.150N} is that thanks to \textit{Gaia} DR2 data a more accurate revision of the cluster star membership was possible, which removed eight objects from their sample. Some additional distant objects were added by following their procedure. Furthermore, also the distance and age scale of the open clusters was completely revised.

The first intention of this paper was to improve our knowledge of the transition radius by adding 14 open clusters to the data set by \citet{2016A&A...585A.150N}. These are located in a somewhat outer Galactic region (\rgc\,$\sim$\,10-12\,kpc) and were analysed with a photometric method introduced by \citet{2010A&A...514A..81P}. During preparation of this paper, another large data set of open cluster metallicities was published \citep{2020AJ....159..199D} that allowed us to tackle the other mentioned issues as well, in particular the age dependence of the radial \feh\ gradient and the influence of radial migration. 

This paper is arranged as follows: In Sect. 2 and 3, we describe the photometric observations and analysis of 14 open clusters. In Sect. 4, we compile spectroscopic metallicities of open clusters, and in Sect. 5, we discuss the Galactic distribution of the samples, the age dependence of the metallicity gradients, and provide an estimate of the radial migration rate. Finally, Sect. 6 summarizes this paper.

\section{Observations and data reduction}

The observations of 14 open clusters were carried out at the San Pedro Martir Observatory (SPMO) during photometric nights in
the years 2002--2019 using the 0.84-m (f/15) Ritchey--Chretien telescope, which is equipped with the Mexman filter wheel and the ESOPO CCD detector. The ESOPO detector, a 2048 $\times$ 2048
13.5-$\mu$m square pixels E2V CCD42-40, has a gain of 1.65 e$^-$/ADU and a readout noise 3.8 e$^-$
at  2 $\times$ 2 binning. The combination of telescope and detector
provides an unvignetted field of view of 7.4$\times$9.3 arcmin$^2$.

Each open cluster was observed at very good seeing conditions (0\farcs6 in long $V$ exposure images) through the Johnson's $UBV$ and the Kron--Cousins' $RI$ filters with short and long exposure times to properly cover both, bright and faint stars in the region. Standard star fields \citep{2009AJ....137.4186L} were observed at the meridian and at about two airmasses to properly determine the atmospheric extinction coefficients. The observation log is provided as supplementary material. Flat fields were taken at the beginning and the end of each night, and bias images were obtained between the cluster observations. 
Data reduction was carried out by one of the authors (RM) with the \textsc{iraf/daophot}\footnote{IRAF is distributed by the National Optical Observatories, operated by the Association of Universities for Research in Astronomy, Inc., under cooperative agreement with the National Science Foundation.} package \citep{1987PASP...99..191S}. 

The standard magnitude for a given filter $\lambda$ was derived using the following relation:
\begin{equation}
M_{\lambda} =  m_{\lambda}-k_{\lambda}X + \eta_{\lambda} C +\zeta_{\lambda},
\end{equation}

where m$_{\lambda}$, k$_{\lambda}$, C and X are the observed instrumental magnitude, extinction coefficients, colour index, and air mass, respectively. M$_{\lambda}$, $\eta_{\lambda}$, and $\zeta_{\lambda}$ are the standard magnitude, transformation coefficient, and photometric zero point, respectively. More details about the data reduction can be found in the papers by \cite{2010RMxAA..46..385A,2015NewA...34..195O,2019JApA...40...33O,2020AN....341...44O}, \cite{2007IAUS..235..331S}, \cite{2010MNRAS.401..621T}.

\section{Open cluster analysis}

\subsection{Membership}

\begin{figure*}
\centering
\resizebox{\hsize}{!}{\includegraphics{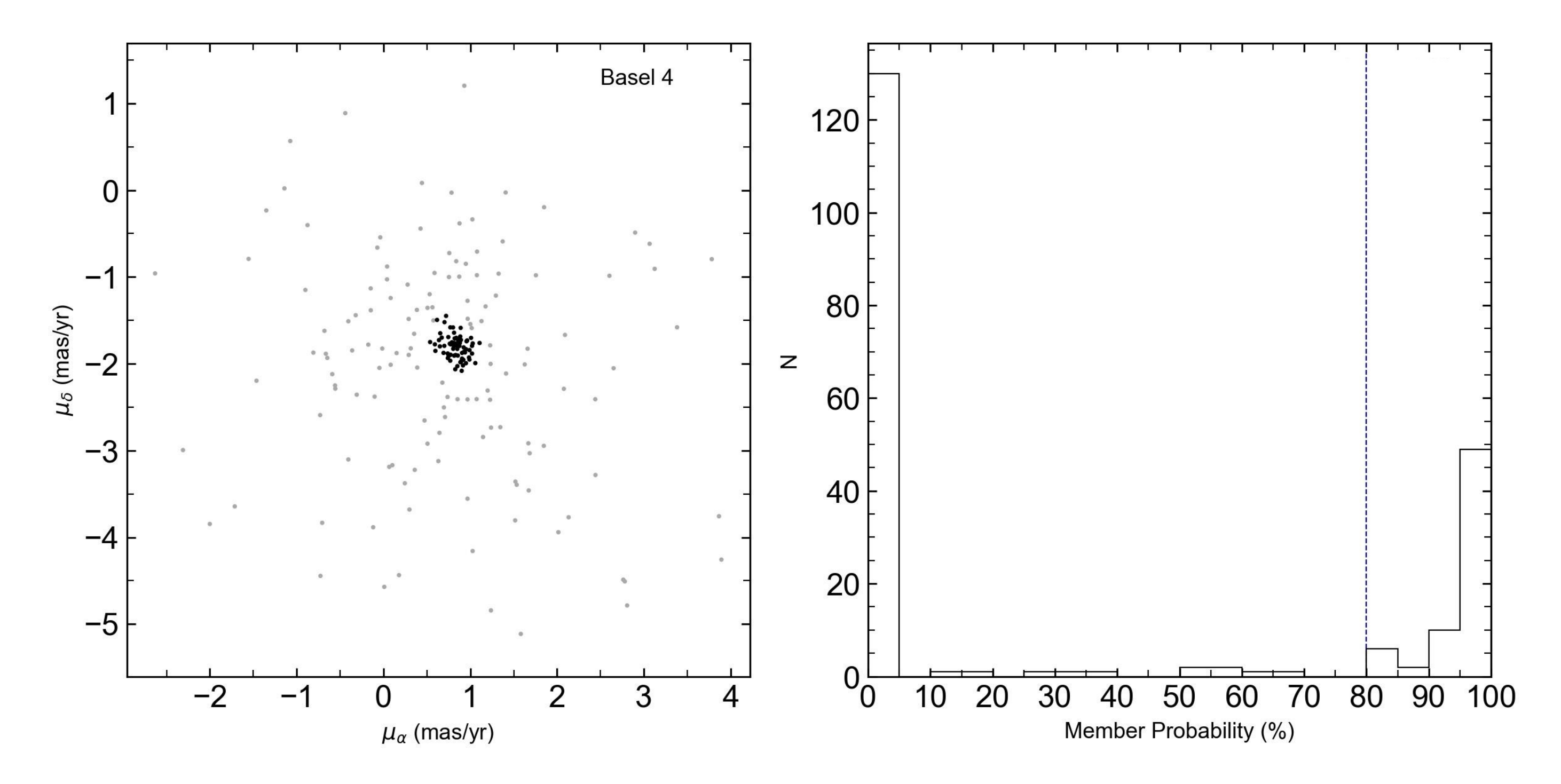}}
\caption{Left-hand panel: the $\mu_{\alpha} - \mu_{\delta}$ distribution of stars in Basel~4. Right-hand panel: the membership probability histogram P(per cent). The stars with $P \geq 80$ per cent (vertical blue dashed line) are adopted as the likely cluster members (filled black dots in the VPD).}
\label{fig:basel4_members}
\end{figure*}

The $UBVRI$ photometric data of the open clusters have been combined with \textit{Gaia} DR2 \citep{2018A&A...616A...1G} proper motion and parallax data to select the most likely cluster members. As a quite typical example, the distribution in $\mu_{\alpha},\mu_{\delta}$ (vector point diagram, VPD) of the cluster stars in Basel~4 is shown in Fig. \ref{fig:basel4_members}.  The potential cluster members show a more concentrated structure, whereas field stars (grey dots) have a more scattered distribution. The membership probabilities P(per cent) of the cluster stars have been determined using a Gaussian Mixture Model \citep[GMM;][]{pedregosa}. The first rise in the P histograms of the clusters are taken as criteria for the cluster membership limit. In the case of Basel~4, stars with P$\geq$80 per cent (vertical dashed line of panel b) are considered to be likely members (filled dots in the VPDs). The same membership procedure has been applied to the remaining open clusters. The GMM \footnote{$P$ is defined as $\Phi_c$ /$\Phi$.  Here $\Phi = \Phi_c + \Phi_f$ is the total probability distribution, where \textit{c} and \textit{f} are subscripts for cluster and field parameters, respectively. The used parameters for the estimation of $\Phi_c$ and $\Phi_f$ are $\mu_{\alpha}$, $\mu_{\delta}$, $\varpi$, $\sigma_{\mu\alpha}$, $\sigma_{\mu\delta}$, $\sigma_\varpi$.} model considers that the distribution of proper motions of the stars in a cluster's region can be represented by two elliptical bivariate Gaussians. The expressions used can be found in the papers by \cite{1998A&AS..133..387B}, \cite{2002A&A...381..464W}, \cite{2012A&A...543A..87S} and \cite{2018MNRAS.481.3887D}.
The median proper motion components, median parallaxes and P(per cent) cut-off values of the probable members are listed in Table\,\ref{tab:clustermembership}. Within the uncertainties the median proper motions and parallax of the likely cluster members are in agreement with the results by \cite{2020A&A...633A..99C}.

% ---------------------------
% ---------------------------

\begin{table*}
	\caption{Membership analysis of the open clusters.  } 
	\label{tab:clustermembership} 
	\begin{center} 
		\begin{tabular}{lrrrrrrrrrr}
			\hline
			Cluster & N & $\mu_{\alpha}$ &$\sigma_{\mu\alpha}$&$\mu_{\delta}$ &$\sigma_{\mu\delta}$& $\varpi$ &$\sigma_\varpi$ & d~(kpc) &$\sigma_{d}$ & P(per cent)\\
			\hline
			Basel 4      &   67  &     0.846  &   0.014  &  $-$1.804  &   0.016  &  0.278  &   0.009  &  3.600  &   0.121  &    80 \\
			Berkeley 35  &   85  &  $-$0.564  &   0.011  &  $-$0.240  &   0.012  &  0.170  &   0.008  &  5.893  &   0.264  &    85 \\
			Berkeley 60  &   94  &  $-$0.666  &   0.013  &  $-$0.720  &   0.011  &  0.230  &   0.008  &  4.347  &   0.153  &    80 \\
			Berkeley 77  &   56  &  $-$0.976  &   0.014  &  $-$0.131  &   0.019  &  0.259  &   0.012  &  3.856  &   0.183  &    80 \\
			Berkeley 104 &   53  &  $-$2.399  &   0.025  &     0.097  &   0.020  &  0.191  &   0.011  &  5.244  &   0.307  &    80 \\
			Haffner 4    &   77  &  $-$0.406  &   0.021  &     0.927  &   0.020  &  0.209  &   0.010  &  4.786  &   0.222  &    80 \\
			King 15      &   67  &  $-$2.334  &   0.020  &  $-$0.823  &   0.017  &  0.308  &   0.010  &  3.252  &   0.101  &    70 \\
			King 23      &   54  &  $-$0.444  &   0.017  &  $-$0.895  &   0.013  &  0.272  &   0.008  &  3.683  &   0.110  &    85 \\
			NGC 1857     &  105  &     0.560  &   0.011  &  $-$1.307  &   0.010  &  0.339  &   0.006  &  2.952  &   0.053  &    80 \\
			NGC 2186     &  106  &     0.398  &   0.014  &  $-$1.977  &   0.025  &  0.409  &   0.009  &  2.447  &   0.052  &    75 \\
			NGC 2236     &  226  &  $-$0.746  &   0.012  &     0.017  &   0.011  &  0.357  &   0.005  &  2.799  &   0.042  &    70 \\
			NGC 2259     &  126  &  $-$0.247  &   0.018  &  $-$1.117  &   0.017  &  0.310  &   0.012  &  3.230  &   0.129  &    70 \\
			NGC 2304     &  102  &  $-$0.029  &   0.016  &  $-$1.548  &   0.020  &  0.198  &   0.008  &  5.063  &   0.199  &    75 \\
			NGC 2383     &  264  &  $-$1.642  &   0.007  &     1.899  &   0.008  &  0.308  &   0.006  &  3.252  &   0.064  &    70 \\
			\hline
		\end{tabular}
	\end{center}
	\flushleft
	\medskip
	\textit{Notes}. Gaia DR2 proper motion components (mas~yr$^{-1}$), parallax (mas), and distances (kpc) of the likely members in the programme clusters. The chosen lower limit of membership probabilities, P(per cent) are given in last column.
\end{table*}

\subsection{Open cluster parameters}
\label{dg_method}

The used open cluster analysis method was developed by \citet{2010A&A...514A..81P}, and was applied with some modifications to a larger sample by \citet{2013A&A...557A..10N}. It uses photometric data that are transformed to effective temperatures and luminosities in an iterative way. These are compared to zero-age main-sequence (ZAMS) normalized isochrones (differential grids, DG) to obtain the open cluster parameters, including metallicity. For more details we refer to the two references above, but we want to emphasize here the main advantages of the method: it relies on the complete cluster MS and allows to incorporate many photometric systems and colours to improve the reliability of the results. 

\citet{2013A&A...557A..10N} have shown that the results for metallicity are independent of the used evolutionary models, the resulting $Z$ values just need to be correctly transformed into the corresponding \feh\ values, the most common used indication of metallicity. Though we note that in case of non-solar scaled abundances, the calibrated \feh\ value might deviate from the actual iron abundance. 

For this study, we calculated new differential grids, because the previously used ones are available only in quite large steps in age ($\Delta$ log\,t = 0.2) and luminosity ($\Delta\,L/\rm{L}_{\odot} = 0.3$) and rely on an already outdated solar abundance ($Z$ = 0.02). We adopt the version 1.2S of the PARSEC tracks \citep{2012MNRAS.427..127B}\footnote{http://stev.oapd.inaf.it/cgi-bin/cmd}, which were computed for a scaled-solar composition with $Z_\odot$ = 0.0152. The ZAMS was constructed using the to the solar value closest available track (Z = 0.014) and isochrones were obtained for seven metallicities ($Z$=0.004, 0.006, 0.010, 0.014, 0.020, 0.030, and 0.040) with age steps    
$\Delta$ log\,t = 0.05. Finally, the differential grids were derived in luminosity steps of 0.1\,dex from $L/\rm{L}_{\odot} = 0.0$ up to the end of the MS. 
 
Some additional improvements in deriving effective temperatures and luminosities of the cluster stars as input to the analysis method rely on the work by \citet{2013ApJS..208....9P} and their updated Table 5\footnote{https://www.pas.rochester.edu/\~{}emamajek [Version 2019.3.22]}. We adopt their results for the bolometric correction instead of the one by \citet{1996ApJ...469..355F}. Furthermore, their photometric colours of the dwarfs effective temperature sequence were used to adjust possible temperature offsets in the compiled photometric temperature calibrations. We noticed that the calibration for the $(V-R)_C$ colour by \citet{2013A&A...557A..10N} resulted in too hot temperatures because of a sign error in the transformation between the different $(V-R)$ systems. The correct temperature calibration to match also the scale by \citet{2013ApJS..208....9P} is 
\begin{equation}
\theta_\mathrm{eff} = 0.536(3) + 0.947(13) (V-R)_{C}.
\end{equation}

The effective temperatures based on the other colours agree very well with the temperature sequence by \citet{2013ApJS..208....9P} at a 0.01-mag level. Finally, we added their temperature sequence for \textit{Gaia} photometry ($G_{BP}-G_{RP}$); thus, the derived mean effective temperatures are based on up to five colour indices: using our photometric data, 2MASS \citep{2006AJ....131.1163S} or UKIDSS \citep{2008MNRAS.391..136L} $K$ magnitudes, and the \textit{Gaia} colour index. The latter colour was dereddened using extinction coefficients as a function of interstellar absorption $A_V$ and the colour ($G_{BP}-G_{RP}$) as suggested by \citet{2018A&A...616A..10G}. For the total-to-selective extinction ratio, we adopt a standard value $R_V = 3.1$ to avoid an additional free parameter in the open cluster analysis.

The comparison of the temperature scales based on the individual colours 
and the colour--colour transformations between \textit{Gaia} and $VRI$ data \citep{2018A&A...616A...4E} allow to evaluate possible offsets of our photometric data (see Table~\ref{tab:offsets}). We noticed a good agreement between these two approaches for $(V-I)$, but found a disagreement for $(V-R)$, depending on the cluster reddening. The difference increases up to about 0.08\,mag for the highest reddened cluster (Berkeley~60). This might be related to the colour excess ratio $E(V-R)/E(B-V)$ given by \citet{1998A&A...333..231B}, but also to reddening or metallicity effects in the colour--colour transformations. We note that $(V-I)$ shows none or little blanketing effects compared to $(V-R)$ \citep[see e.g.][]{1996A&A...313..873A}. However, using the offsets based on the temperature scale, we noticed a good agreement if fitting isochrones to the individual CMDs. 

\begin{figure*}
\centering
\resizebox{\hsize}{!}{\includegraphics{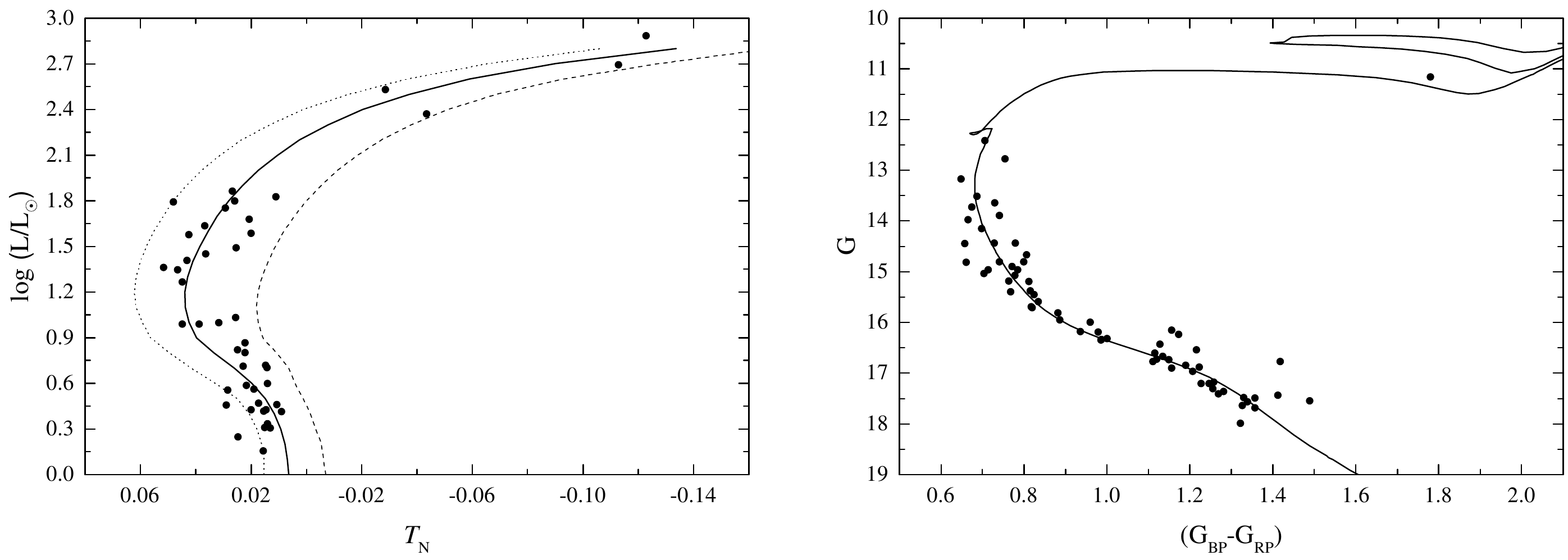}}
\caption{Result for the open cluster Basel~4. The left-hand panel shows the best fit based on the differential grids. The thick line represents the isochrone for $\log\,t=8.20$ and Z=0.01, the dotted line for Z=0.006, and the dashed line for Z=0.02. $T_N$ is the temperature difference in dex between the star and the ZAMS at solar metallicity using the mean temperature based on up to five colours. The right-hand panel shows the \textit{Gaia} CMD and an isochrone with the parameters given in Table\,\ref{tab:clusterresults}.}
\label{fig:basel4}
\end{figure*}

As starting values for the open cluster analysis we used a mix of \textit{Gaia} based distances by \citet{2020A&A...633A..99C}, the parameter estimates by \citet{2020A&A...640A...1C}, and our extinction values based on the ZAMS fitting. For the least known parameter metallicity we initially adopt the solar value. The starting values were altered in an iterative procedure until the best fit with the derived differential grids was found (lowest $\sigma$ over the complete luminosity range). The choice of the starting values has no influence on the final results, but reduces in ideal circumstances the number of iterations. A consistency check of the parameter results was performed by applying PARSEC isochrones to all available CMDs. The derived $Z$ values were transformed to \feh\ using the relation $\log(Z/X)-\log(Z/X)_{\odot}$, with $(Z/X)_{\odot}=0.0207$ and $Y=0.2485+1.78Z$ \citep{2012MNRAS.427..127B}. All results for the open clusters are listed in Table~\ref{tab:clusterresults}, and as an example we present the fits for Basel~4 in Fig.~\ref{fig:basel4}. For the remaining open clusters we refer to the Figs. \ref{fig:be35} - \ref{fig:n2383}. For most objects, the derived parameters also provide a reasonable fit of the isochrones to the CMDs. However, differential reddening and also the fixed $R_V$ value in the analysis might lead to apparent offsets in some colours (see e.g. Berkeley~60 in Fig. \ref{fig:be60}).

The applied method identifies the most reasonable result for the Z-parameter with the lowest scatter over the complete luminosity range; thus, the errors of the remaining cluster parameters (reddening, distance, and age) are not straightforward to indicate. Though, based on the iteration process we conclude that the errors are generally within two iteration steps for the age and distance and within three iteration steps for the reddening; thus, 0.1\,dex for the age, 0.1\,mag for the distance modulus, and 0.03\,mag for the reddening. 

Spectroscopic metallicities were recently presented by \citet{2020AJ....159..199D} for three programme clusters using the SDSS/APOGEE survey. They quote $\feh = -0.16, -0.14, and -0.12$\,dex for Haffner~4, NGC~1857, and NGC~2304, respectively. The results are in good agreement to our estimates, but on average a few hundreds of dex (0.02) more metal deficient. Eight more clusters in this sample were previously analysed with the DG method, showing an offset of $0.01 \pm 0.06$\,dex (own $-$ APOGEE). Thus, indicating that the change in the analysis used in this study has not altered the previous metallicity scale. This confirms the conclusion by \citet{2013A&A...557A..10N}. Using all clusters in common we derive an offset to the APOGEE results of $0.01 \pm 0.05$\,dex. Prior to this study, almost 70 open clusters were analysed with this method and \citet{2016A&A...585A.150N} noticed an almost identical accuracy and precision in comparison with their spectroscopic sample.

% ---------------------------
\begin{table*}
\caption{Derived parameters of the open clusters.  } 
\label{tab:clusterresults} 
\begin{center} 
\begin{tabular}{l c c c c c r} 
\hline
Cluster & $\log$\,$t$   & $E(B-V)$  & $(m-M)_0$   & $Z$  & \feh   & \rgc \\ 
        & (dex) & (mag) & (mag) &   &  (dex) & (kpc) \\
\hline 
Basel~4 & 8.20 & 0.55 & 12.50 & 0.010(3) & $-$0.18(14) & 11.16 \\
Berkeley~35 & 9.00 & 0.13 & 13.50 & 0.011(2) & $-$0.13(08) & 12.50 \\
Berkeley~60 & 8.45 & 0.86 & 12.85 & 0.013(3) & $-$0.06(11) & 10.32 \\
Berkeley~77 & 8.95 & 0.10 & 12.80 & 0.012(3) & $-$0.09(12) & 11.04 \\
Berkeley~104 & 8.90 & 0.52 & 13.10 & 0.013(4) & $-$0.06(14) & 10.60 \\
Haffner~4 & 8.65 & 0.45 & 13.00 & 0.011(3) & $-$0.13(13) & 11.06 \\
King~15 & 8.70 & 0.54 & 12.35 & 0.013(4) & $-$0.06(14) & 9.84 \\
King~23 & 9.25 & 0.11 & 12.55 & 0.010(2) & $-$0.18(09) & 10.74 \\
NGC~1857 & 8.45 & 0.48 & 12.20 & 0.012(4) & $-$0.09(16) & 10.71 \\
NGC~2186 & 7.95 & 0.31 & 11.75 & 0.012(2) & $-$0.09(08) & 10.08 \\
NGC~2236 & 8.90 & 0.48 & 12.10 & 0.013(3) & $-$0.06(11) & 10.45 \\
NGC~2259 & 8.60 & 0.57 & 12.30 & 0.011(3) & $-$0.13(13) & 10.73 \\
NGC~2304 & 9.00 & 0.05 & 12.85 & 0.011(3) & $-$0.13(13) & 11.56 \\
NGC~2383 & 8.50 & 0.30 & 12.50 & 0.013(3) & $-$0.06(11) & 10.14 \\
\hline 
\end{tabular}
\end{center}
\flushleft
\medskip
\textit{Notes}. For $Z$ and \feh, the errors of the last significant digits are given in parenthesis. The errors of the other cluster parameters are discussed in the text. These are 0.1\,dex for the age, 0.1\,mag for distance modulus, and 0.03\,mag for $E(B-V)$. 
\end{table*}
% ---------------------------

% ---------------------------
% ---------------------------

\section{Spectroscopic metallicities}

\subsection{Data compilation}
\label{datacompilation}

In addition to the new metallicity estimates for our sample clusters, we made use of the spectroscopic data listed by \citet{2016A&A...585A.150N}. The list was updated accordingly with additional data, but including only objects beyond the solar circle ($R_{\rm GC} > 9~\rm{kpc}$) as the solar vicinity is already very well covered. We noticed four clusters in three references: Berkeley~25 and NGC~2243 by \citet{2017A&A...603A...2M}, NGC~1907 and NGC~7245 by \citet{2017MNRAS.470.4363C}, and NGC~2243 by \citet{2019A&A...623A..80C}. The results for Berkeley~25 and few stars in NGC~2243 by \citet{2017A&A...603A...2M} belong to the lower quality (LQ) data sample as defined by \citet{2016A&A...585A.150N} due to the signal-to-noise ratio, all others were merged with the high-quality (HQ) sample. Furthermore, the study by \citet{2019A&A...623A..80C} used APOGEE and GALAH data, but we only adopt their results based on the latter survey for NGC~2243. 

\citet{2020AJ....159..199D} presented an investigation of open clusters using the most recent SDSS/APOGEE data release (DR16) that we adopt instead. They introduced a reliability flag of the results based on visual inspection of the stars' position in the CMD, and we restrict the total sample of 128 open clusters to the high-quality sample of 71 objects. However, we noticed that one of the outermost clusters (Saurer~1) was flagged as potentially unreliable. We checked the position of the star in the CMD, which although is  separated from the main bulk of stars, it lies very close to the corresponding isochrone in all available colours (see Fig. \ref{fig:sau1}). We thus keep the result for this cluster in the sample, because its metallicity is very close to the data in the HQ sample as well. The whole APOGEE data set was kept as an independent one, because of its size and homogeneity. However, based on the spectral resolution of APOGEE (R$\sim$22\,500), it would belong to the LQ sample by \citet{2016A&A...585A.150N}. 

The membership of the cluster stars in the APOGEE sample was already verified using \textit{Gaia} DR2 \citep{2018A&A...616A...1G}. In the HQ and LQ sample, several cluster results are based on single or few stars, thus it is important to check the cluster membership in these samples as well. We used the membership analysis by \citet{2020A&A...633A..99C} and \textit{Gaia} data for not covered objects to conclude about membership based on proper motion, parallax, and radial velocities. This removed six clusters in the HQ sample (Melotte~71, NGC~1545, NGC~1901, NGC~1977, NGC~2266, and NGC~2335) and two objects in the LQ sample (Berkeley~21 and Berkeley~75).  

The total sample with spectroscopic metallicities includes 138 open clusters. We compare the individual data sets and derive the following 2$\sigma$ clipped offsets: $0.05 \pm 0.05$ (HQ$-$APOGEE, 24 objects), $-0.01 \pm 0.07$ (LQ$-$APOGEE, 15 objects), and $0.03 \pm 0.08$ (HQ$-$LQ, 31 objects). The two data sets based on medium resolution agree very well, but the HQ data are more metal rich than the APOGEE results. Though, the offset between the HQ and APOGEE data reduces to about 0.02\,dex if using the number of measured stars as a weight. The APOGEE DR16 data include a zero-point shift to force solar metallicity stars in the solar neighbourhood to have a mean [X/M]=0 \citep{2020AJ....160..120J}. A similar comparison of clusters close to the solar circle (7.5\,kpc $<$ \rgc\ < 8.5\,kpc) in all three samples shows a mean metallicity in the range  $0.00 \pm 0.01$\,dex. Thus, for the following analysis we do not apply any offsets.

\subsection{Distance and age scale}

\citet{2020AJ....159..199D} nicely demonstrated the effect of the distance scale on the analysis of the metallicity gradient. The cluster age is crucial as well, but generally even worse defined than the distance as shown e.g. by \citet{2015A&A...582A..19N}. Thus, it is important to adopt proper cluster parameters. For small samples it is certainly possible to derive them oneself or check available parameters in a homogeneous way, but for larger samples one has to rely on literature and to decide about the most reliable source. 

One can either pick results from numerous individual studies at the expense of homogeneity, or adopt large-scale studies that derived cluster parameters in an automatic way. 
There are several automatic approaches available \citep[see e.g.][]{2015A&A...582A..19N}, a recent one by \citet{2020A&A...640A...1C}, who applied an artificial neural network (ANN) on 2D histograms of \textit{Gaia} CMDs using parallax as an additional parameter. Such approaches generally cannot identify individual special features or discrepancies in the parametrization that are noticeable only in a detailed by-eye examination, but a visual inspection of the results as performed e.g. by the latter reference above is usually included.

For this study, we adopt a combination of the two possibilities mentioned above. We use the DG results based on a multicolour analysis \citep[][and this paper]{2013A&A...557A..10N,2016A&A...585A.150N} as primary source for 31 clusters in the spectroscopic sample, followed by results from The Bologna Open Clusters
Chemical Evolution project \citep[BOCCE, see e.g.][]{2006AJ....131.1544B} for 21 clusters, and \citet{2019A&A...623A.108B}, who used \textit{Gaia} data of 269 selected objects that were analysed with an automated Bayesian tool to fit isochrones. This work covers additional 24 clusters of our sample. The two studies were adopted by \citet{2020A&A...640A...1C} for the training sample of the ANN. Finally, for the remaining objects we rely on the results by the last reference.  

Three clusters from the APOGEE sample are not covered by the selected references (Berkeley~43, FSR~394, and Saurer~1). The first object was flagged by \citet{2020A&A...640A...1C} as too red, the others are not included (or were rejected) by this survey. We inspected the \textit{Gaia} CMDs and conclude that FSR~394 might be an old distant cluster and Berkeley~43 a highly reddened cluster of intermediate age. However, the large scatter in the CMDs do not allow careful isochrone fits. We therefore rejected the two objects from the sample. Saurer~1, on the other hand, shows a well-defined giant branch and we initially adopt the result by \citet{2003MNRAS.346...18C}, who present deep $VI$ photometry of the cluster and derived an age of $\log t = 9.7$.

Finally, we verified the parameters of the 16 most distant ($\gtrsim 5$\,kpc) clusters, corresponding to galactocentric distances of $R_{\rm GC} \gtrsim 12.5$\,kpc using \textit{Gaia} CMDs of likely proper motion members and isochrones for metallicities that roughly correspond to the available spectroscopic measurements.

\begin{figure*}
\centering
\resizebox{\hsize}{!}{\includegraphics{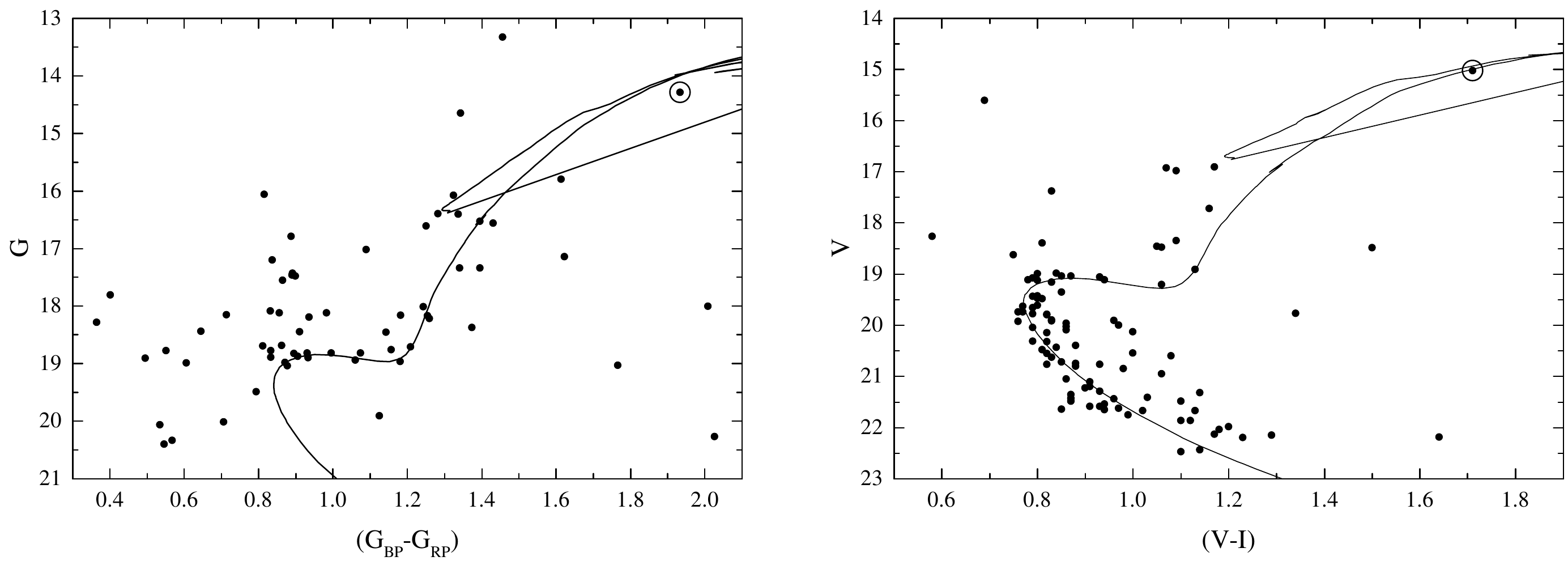}}
\caption{Result for the open cluster Saurer~1. The left-hand panel shows the \textit{Gaia} CMD of proper motion members and the right-hand panel shows the $VI$ data by \citet{2003MNRAS.346...18C} restricted to the innermost cluster region. The overplotted isochrones correspond to the parameters listed in Table \ref{tab:dist-clusters}. The position of the star included in the spectroscopic study by \citet{2020AJ....159..199D} is marked with a circle.}
\label{fig:sau1}
\end{figure*}

For some objects, such as Saurer~1 or Berkeley~29 (see Figs. \ref{fig:sau1} and \ref{fig:be29}), \textit{Gaia} photometry is not deep enough to properly characterize the cluster. In such cases, we queried for additional deeper photometry. Table \ref{tab:dist-clusters} lists the initial and the adopted parameters after revision. In general, we notice a good agreement, though there are some objects for which the distance differs by about 20 per cent (Berkeley~19, Berkeley~21, Berkeley~22, and Berkeley~25). An excerpt of the physical parameters of the complete open cluster sample is shown in Table \ref{tab:physparameter}; the full table is available at the CDS.

% ---------------------------
\begin{table*}
\caption{Parameters of the most distant sample clusters.  } 
\label{tab:dist-clusters} 
\begin{center} 
\begin{tabular}{lcccccccc} 
\hline
Cluster & $\log t$   & d (pc)  & $E(B-V)$   & source & $\log t$ & d (pc)  & $E(B-V)$ & \feh  \\ 
   & \multicolumn{3}{c}{Initial parameters} & &  \multicolumn{3}{c}{Revised parameters} & \\
\hline 
Berkeley 18 & 9.64 & 5632 & 0.49 & 1 & 9.50 & 6310 & 0.55 & -0.40 \\
Berkeley 19 & 9.34 & 6568 & 0.27 & 1 & 9.40 & 5495 & 0.42 & -0.30 \\
Berkeley 20 & 9.76 & 8710 & 0.13 & 2 & 9.75 & 8710 & 0.17 & -0.40 \\
Berkeley 21 & 9.34 & 5012 & 0.78 & 2 & 9.30 & 6026 & 0.70 & -0.30 \\
Berkeley 22 & 9.38 & 5754 & 0.64 & 2 & 9.40 & 6918 & 0.63 & -0.30 \\
Berkeley 25 & 9.39 & 6780 & 0.35 & 1 & 9.60 & 7943 & 0.35 & -0.25 \\
Berkeley 29 & 9.57 & 13\,183 & 0.12 & 2 & 9.60 & 13\,183 & 0.09 & -0.40 \\
Berkeley 31 & 9.46 & 7586 & 0.19 & 2 & 9.45 & 7586 & 0.20 & -0.40 \\
Berkeley 33 & 8.37 & 5851 & 0.56 & 1 & 8.50 & 5754 & 0.61 & -0.30 \\
Berkeley 73 & 9.15 & 6158 & 0.22 & 1 & 9.20 & 7244 & 0.27 & -0.20 \\
Czernik 30 & 9.46 & 6647 & 0.20 & 1 & 9.40 & 7244 & 0.32 & -0.40 \\
IC 166 & 9.12 & 5285 & 0.80 & 1 & 9.10 & 5248 & 0.83 & -0.10 \\
King 2 & 9.61 & 6760 & 0.31 & 1 & 9.70 & 6607 & 0.40 & -0.40 \\
NGC 1798 & 9.22 & 5124 & 0.36 & 1 & 9.20 & 4571 & 0.47 & -0.30 \\
Saurer 1 & 9.70 & 13\,183 & 0.14 & 3 & 9.70 & 13\,183 & 0.17 & -0.40 \\
Tombaugh 2 & 9.20 & 7943 & 0.34 & 2 & 9.30 & 9120 & 0.30 & -0.30 \\
\hline 
\end{tabular}
\end{center}
\flushleft
\medskip
\textit{Notes}. The sources of the initial parameters are as follows: (1) \citet{2020A&A...640A...1C}, (2) BOCCE, and (3) \citet{2003MNRAS.346...18C}. The last column lists the adopted metallicity for the corresponding isochrones.  
\end{table*}
% ---------------------------

% ---------------------------
% ---------------------------

% ---------------------------
\begin{table*}
\caption{Physical parameters of the 136 open clusters with spectroscopic metallicities -- full table available as supplementary material at the CDS.  } 
\label{tab:physparameter} 
\begin{center} 
\begin{tabular}{lcccccccccc} 
\hline
Cluster & $l$   & $b$  & $\log t$   & Dist. & \feh & $\sigma$\feh  & No. stars / meas. & Source & \rgc\  \\ 
        & (deg) & (deg) &  & (kpc) & (deg) & (deg) & & & (kpc) \\
\hline 
NGC 6583 & 9.283 & $-$2.534 & 9.08 & 2.05 & 0.37 & 0.04 & 2 / 2  & 2 & 5.99  \\
NGC 6494 & 9.894 & 2.834 & 8.68 & 0.66 & $-$0.04 & 0.08 & 3 / 6 & 2 & 7.35 \\
Blanco 1 & 15.572 & $-$79.261 & 7.98 & 0.24 & 0.04 & 0.07 & 6 / 6  & 2 & 7.96 \\
Ruprecht 147 & 21.012 & $-$12.816 & 9.48 & 0.32 & 0.12 & 0.03 & 27  & 1 & 7.71 \\
NGC 6705 & 27.307 & $-$2.776 & 8.40 & 2.09 & 0.12 & 0.04 & 12  & 1 & 6.22 \\
... & ... & ... & ... & ... & ... & ... & ...  & ... & ... \\
\hline 
\end{tabular}
\end{center}
\flushleft
\medskip
\textit{Notes}. The number of stars and measurements is given for the HQ and LQ samples. The source of the metallicity is as follows: (1) \citet{2020AJ....159..199D}; (2) HQ sample; (3) LQ sample. 
\end{table*}
% ---------------------------

% ---------------------------
\begin{table*}
\caption{Orbital parameters of the 146 open cluster with spectroscopic and photometric metallicities -- full table available as supplementary material} at the CDS. 
\label{tab:orbitparameter} 
\begin{center} 
\begin{tabular}{lccccccccc} 
\hline
Cluster & $RV$   & $\sigma RV$  & No. stars   & \rgc\ & $R_{\rm apo}$ & $R_{\rm peri}$ & \rguide & ecc & $z_{\rm max}$  \\ 
        & (km\,s$^{-1}$) & (km\,s$^{-1}$) &  & (kpc) & (kpc) & (kpc) & (kpc) & & (kpc) \\
\hline 
NGC 6583 & $-$1.43 & 0.49 & 24 & 5.99 & 6.75 & 5.97 & 6.34 & 0.06 & 0.09 \\
NGC 6494 & $-$8.00 & 0.13 & 12 & 7.35 & 7.67 & 7.35 & 7.51 & 0.02 & 0.07 \\
Blanco 1 & 6.23 & 0.07 & 172 & 7.96 & 8.49 & 7.96 & 8.16 & 0.03 & 0.22 \\
Ruprecht 147 & 42.18 & 0.38 & 99 & 7.71 & 8.95 & 6.26 & 7.37 & 0.18 & 0.26 \\
NGC 6705 & 34.49 & 0.27 & 357 & 6.22 & 6.53 & 5.51 & 5.98 & 0.08 & 0.09 \\
... & ... & ... & ... & ... & ... & ... & ... & ... & ... \\
\hline 
\end{tabular}
\end{center}  
\end{table*}
% ---------------------------

\subsection{Orbital parameters}
\label{sect:orbit}
Recently, \citet{2021A&A...647A..19T} presented orbital parameters for almost 1400 open clusters. Though, they adopted a different solar distance and also the cluster parameters differ to our scale. We thus repeated their calculations using our adopted cluster parameters, a solar distance of \rgc\ = 8.0\,kpc, the proper motions by \citet{2020A&A...633A..99C} and the compiled mean radial velocities by \citet{2021A&A...647A..19T}. 

Three objects are not covered by the latter reference  (Berkeley~104, Ruprecht~4, and Saurer~1) and for six clusters we noticed errors in the radial velocity in the range 8 - 46\,km\,s$^{-1}$. For these objects we queried the additional literature to adopt the missing or probable more reliable velocities. For the cluster NGC~2194, for example, we adopt +7.5\,$\pm$\,0.8\,km\,s$^{-1}$ by \citet{2011AJ....142...59J} based on seven objects, while \citet{2021A&A...647A..19T} list +38.7\,$\pm$\,8.1\,km\,s$^{-1}$ based on nine stars. Actually, also \textit{Gaia} data suggest the lower velocity, because of a clustering around +9\,km\,s$^{-1}$ of the innermost objects. All of these stars show a proper motion membership probability of 1.0 according to \citet{2020A&A...633A..99C}; thus, a higher threshold than 0.4 adopted by \citet{2021A&A...647A..19T} or some limitation of the radius is required to pinpoint the cluster velocities.  It is out of scope of this work to re-evaluate all radial velocities, we therefore stress that this is certainly an invaluable data set for  statistical analyses, but might lead to wrong conclusions for probably some individual objects. Only one object remained without a radial velocity estimate, Berkeley~104 from our photometric sample.

The orbital calculations were performed with the \textsc{python} package \textsc{galpy}\footnote{http://github.com/jobovy/galpy [version 1.6.0]} by \citet{2015ApJS..216...29B} using the Milky-Way-like potential \texttt{MWPotential2014}, a local rotation velocity of 239\,km\,s$^{-1}$ \citep{2011AN....332..461B} and the solar motion by \citet{2010MNRAS.403.1829S}. Each cluster was integrated with an integration step of 0.01 up to 500\,Myr at maximum to avoid inaccuracies in the time-dependence of the potential \citep[see discussion by][]{2021A&A...647A..19T}.

We obtained among others the orbit boundary information $R_{\rm apo}$, $R_{\rm peri}$, $z_{\rm max}$, the orbits' eccentricity $e$, and the guiding radii \rguide. The latter is the radius of a circular orbit with angular momentum $L_{z}$. An excerpt of the orbital parameters are listed in Table \ref{tab:orbitparameter}; the full table is available at the CDS. The eccentricities and galactic plane distances ($z_{\rm max}$) of our sample clusters reflect the properties of the Galactic thin disc.

\section{Results and discussion}

\subsection{Radial metallicity distribution}
\label{sect:rmd}

\begin{figure}
\centering
\resizebox{\hsize}{!}{\includegraphics{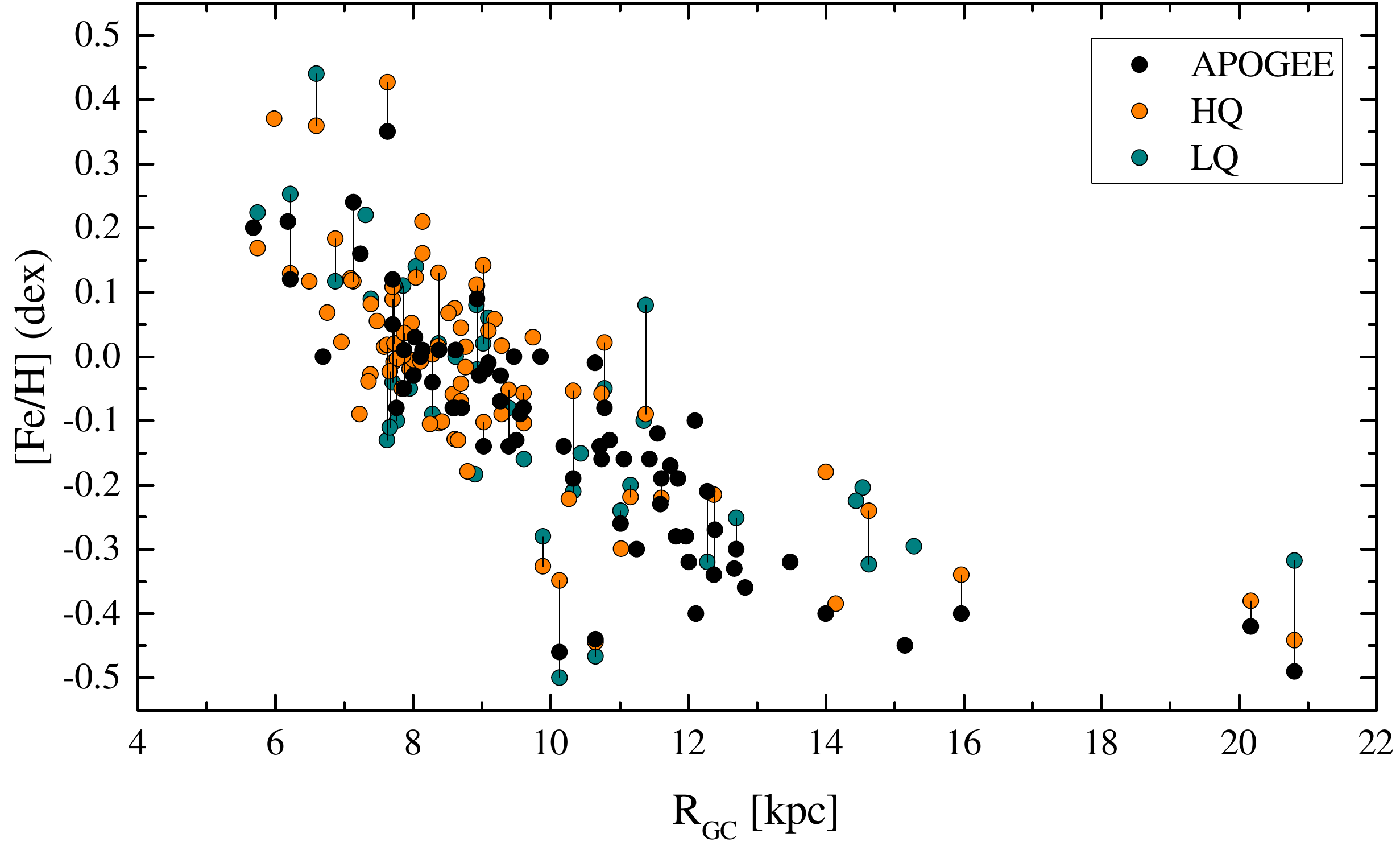}}
\caption{The galactic metallicity distribution based on open clusters with spectroscopic determinations. The connecting lines show objects in common in the three samples. }
\label{fig:specdistribution}
\end{figure}

Figure \ref{fig:specdistribution} shows the spectroscopic metallicities as a function of the galactocentric distance \rgc, adopting 8.0\,kpc as the distance of the Sun from the Galactic Centre for eased comparison with other works \citep[e.g.][]{2016A&A...585A.150N,2020AJ....159..199D}. In total, there are 136 open clusters, 85 and 42 clusters with HQ and LQ data, respectively, and 70 objects covered by the APOGEE survey. 

% ---------------------------
% ---------------------------

For better clarity, we omit the indication of errors in \feh. We note that these are also not easily comparable as the HQ and LQ compilations include a variety of references that already might introduces a larger scatter, depending on the number of individual references and number of stars involved. The mean error in the HQ and LQ samples is 0.05 and 0.07\,dex, respectively, but reach up to 0.13\,dex for even well covered objects (e.g. NGC~752 or NGC~3680 in the LQ sample). The APOGEE sample on the other hand is a homogeneous data set that lists errors in the range of 0.01--0.06\,dex, apparently adopting the internal error (0.01--0.02\,dex) for cluster results based on a single star, while the best covered clusters with 60 or more stars (Melotte~20 and Melotte~22) show an error of 0.05\,dex. Metallicity spreads can be expected due to atomic diffusion for example \citep{2019ApJ...874...97S} 

The radial metallicity distribution of the different samples show a comparable behaviour, but the APOGEE data span a somewhat narrower metallicity range at the different galactocentric distances. This might be related to the homogeneous analysis, but could be a bias due to the target sample as well, as several confirmed under- or overabundant objects are not covered by this survey.

\begin{figure}
\centering
\resizebox{\hsize}{!}{\includegraphics{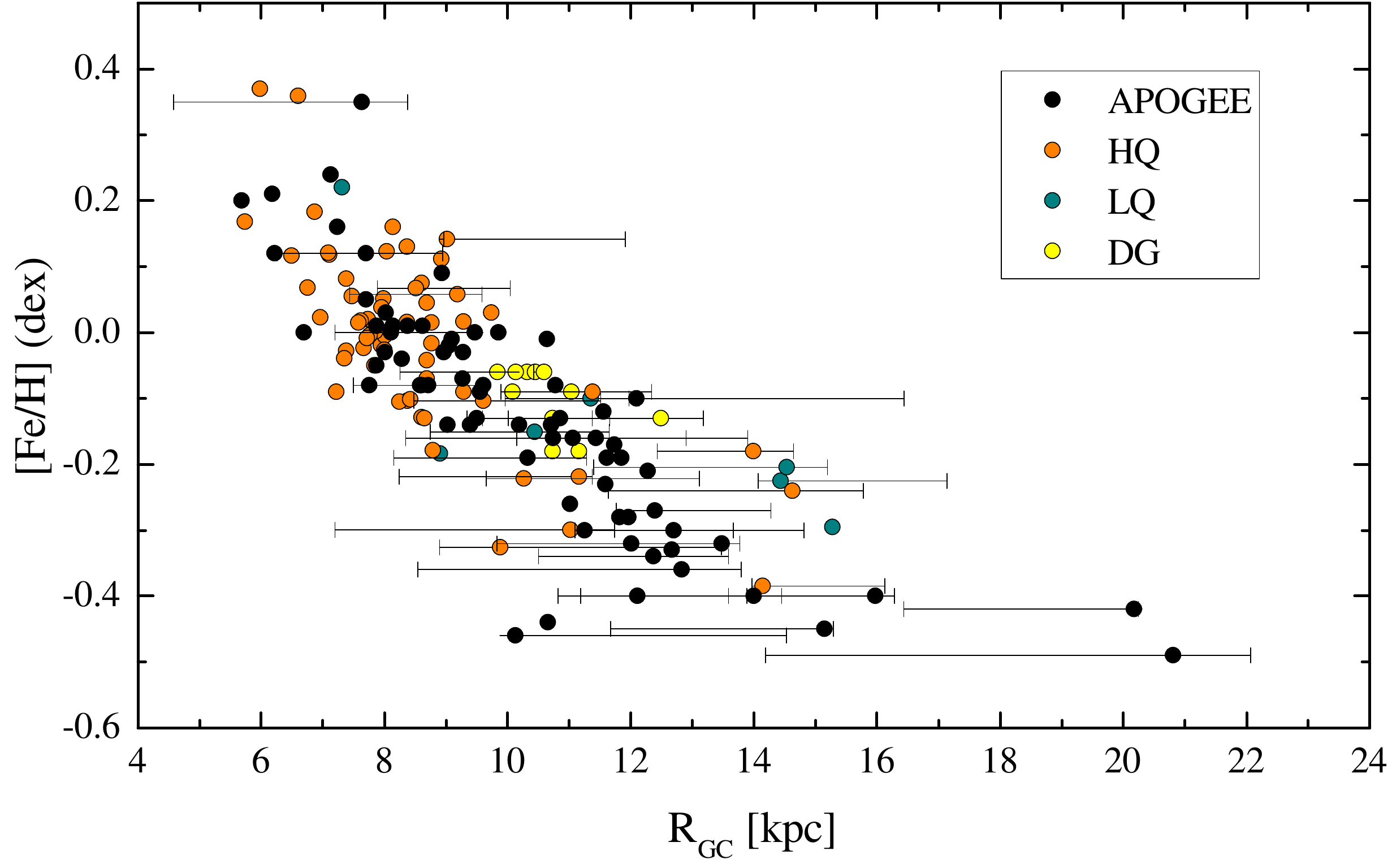}}
\resizebox{\hsize}{!}{\includegraphics{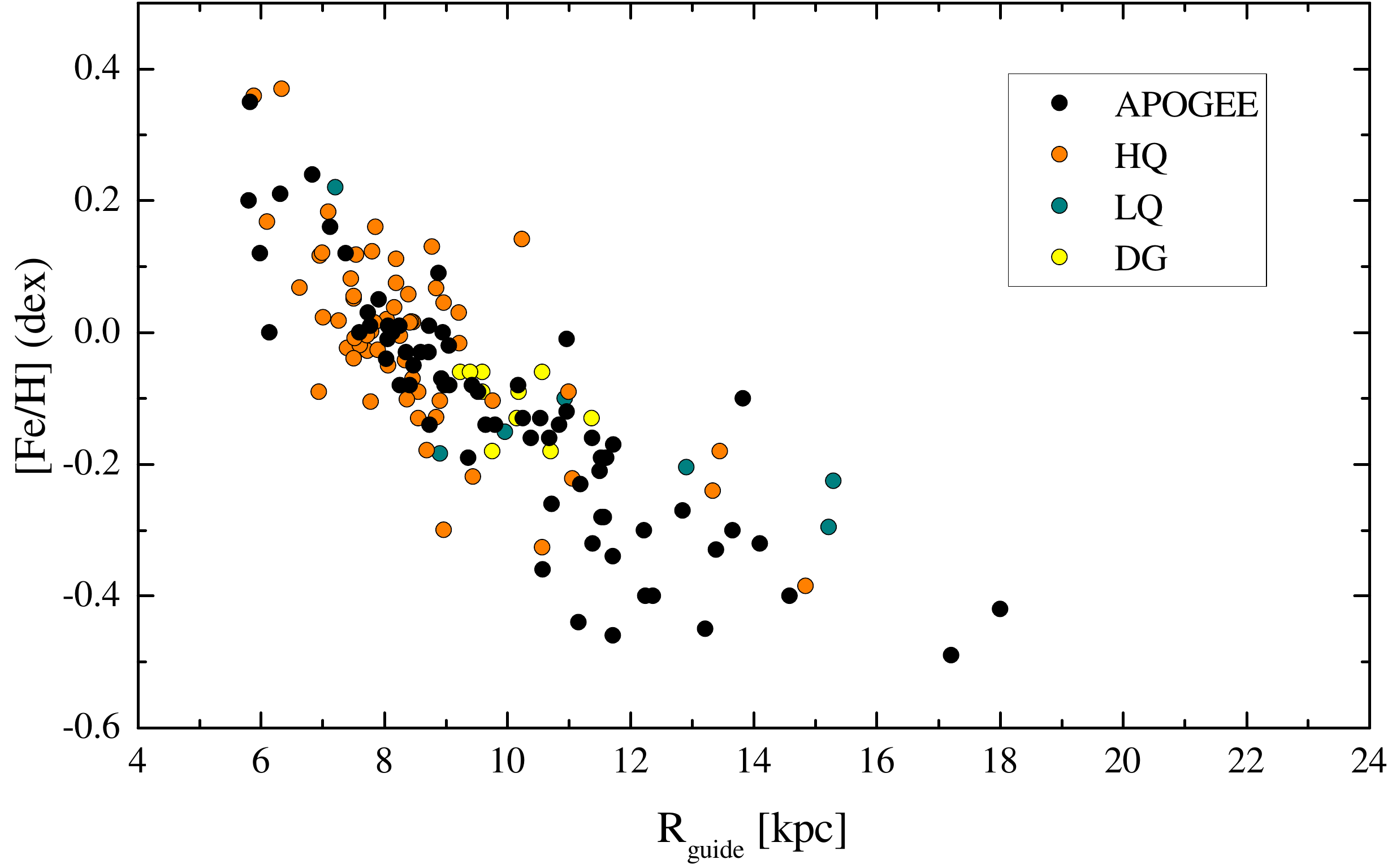}}

\caption{The galactic metallicity distribution based on the merged sample with spectroscopic determinations and the 11 objects presented in Sect. \ref{dg_method} without a spectroscopic measurement. The upper panel shows the distribution for \rgc, also the distance range of $R_{\rm apo}$ and $R_{\rm peri}$ for objects with orbit boundaries larger than 2\,kpc is given. The lower panel presents the metallicity distribution for \rguide.}
\label{fig:merged}
\end{figure}

In the following we merge the individual data sets to a single one by adopting APOGEE data as primary source, thus preferring homogeneity instead of spectral resolution, followed by 59 objects from the HQ compilation and seven clusters with LQ data. 
Figure \ref{fig:merged} shows this merged sample in addition to the 11 objects analysed with the DG method that have no spectroscopic measurement available. The latter nicely match the spectroscopic distribution. 

% ---------------------------
% ---------------------------

Previous works \citep[see][and references therein]{2016A&A...585A.150N} noticed a transition of the metallicity gradient between the inner and outer Galactic disc. While the inner disc shows a steeper gradient, the metallicity distribution of the outer area is almost flat. The above reference adopt \rgc \,$\sim$ 12\,kpc for this transition radius. More recently, \citet{2020AJ....159..199D} used a two-line function fit to derive \rgc \,$\sim$ 13.9\,kpc, which however strongly depends on the adopted distance scale. Though the appearance of a ``knee'' in the metallicity distribution not only depends on the distance scale (note that the most distant objects were verified), also on the covered objects. Compared to \citet{2016A&A...585A.150N} several outer disc clusters that led to a conclusion about a flat gradient are either much closer, more metal deficient or are even rejected (Berkeley~75, see Sect. \ref{datacompilation}) in the present data set. Furthermore, our data set includes nine additional objects more distant than \rgc\,$\sim$\,12\,kpc with spectroscopic metallicities and several photometrically analysed objects (Sect. \ref{dg_method}) out to about \rgc\,$\sim$\,12\,kpc that provide a more clearer picture.  At first glance, Fig. \ref{fig:merged} does not indicate the existence of a gradient change, in particular if the two most distant objects (Berkeley~29 and Saurer~1) are tuned out. 

Using the total sample as presented in Fig. \ref{fig:merged} (without the two objects mentioned above) and a linear regression model that is robust to outliers \citep{1981rost.book.....H} gives a gradient of $-$0.058(4)\dexkpc. The clearly lowest weight was assigned to six objects, most of them already identified  as ``outliers'' by \citet{2016A&A...585A.150N}: Melotte~66, NGC~2243, NGC~6253, NGC~6583, NGC~6791, and Trumpler~5. Repeating the analysis for inner disc objects only (\rgc \,$\leq$\,12\,kpc) results in a gradient of $-$0.063(5) \dexkpc, the exclusion of the six deviating objects gives $-$0.058(5) \dexkpc. We note, however, that the used regression model does only identifies possible outliers in the dependent variable.

\begin{figure}
\centering
\resizebox{\hsize}{!}{\includegraphics{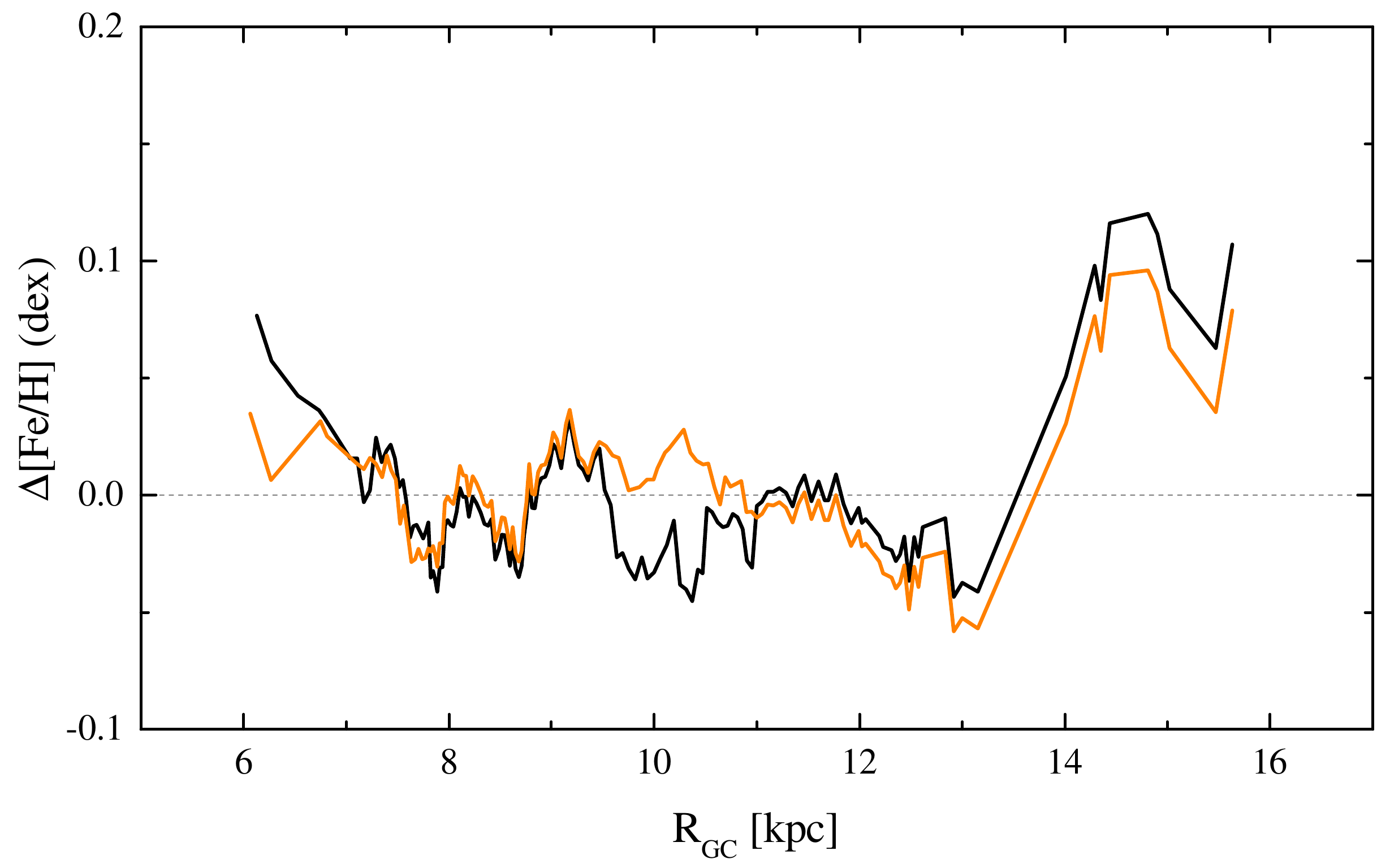}}
\caption{Residuals of the running average for the complete (black line)
and cleaned sample (orange line) after subtracting the respective gradients
of the inner disc objects.}
\label{fig:residuals}
\end{figure}

For the identification of gradient changes, we use a running average on the residuals after subtraction of the gradients of inner disc objects. We grouped the sample either by a constant number of 15 clusters or by a maximum distance range of 1 kpc, whichever criterion was fulfilled first \citep[see][for details]{2016A&A...585A.150N}. Figure \ref{fig:residuals} indicates that the gradient flattens slightly out beyond \rgc\,$\sim$\,13\,kpc, though to a much lower extent as shown e.g. by \citet{2016A&A...585A.150N}. However, the little coverage of objects beyond this region, the split in the metallicity distribution of HQ/LQ and APOGEE data (see Fig. \ref{fig:merged}), and the unequal distribution in age -- most of them are older then $\gtrsim$\,2\,Gyr (see Fig. \ref{fig:agegroups}) -- might result in an incorrect conclusion. Future studies, spectroscopically or even photometrically, are still needed to provide a better coverage of objects more distant than \rgc\,$\sim$\,13\,kpc.

% ---------------------------

\subsection{Age dependence of the radial [Fe/H] gradient}
\label{sect:amr}
An analysis of the overall metallicity gradient using the complete open cluster sample introduces a bias as one mixes populations with different ages. To obtain some idea about the time evolution of the metallicity gradient, one therefore has to restrict the sample to objects with comparable ages as already presented in several previous works \citep[e.g.][to mention just a few]{1998MNRAS.296.1045C,2016A&A...585A.150N,2020AJ....159..199D}. Not only the number of available cluster data increased significantly since the work by \citet{1998MNRAS.296.1045C} who used 37 objects in total, but certainly also the data quality the analysis is based on.

Our sample is large enough to divide it into age groups that still include a sufficient number of objects and span a wide range of galactocentric distances. We define eight consecutive and overlapping age groups to trace variations in more detail: $<$\,0.4\,a, 0.3\,$\leq$\,a\,$<$0.7, 0.4\,$\leq$\,a\,$<$\,1.0, 0.7\,$\leq$\,a\,$<$\,1.5, 1.0\,$\leq$\,a\,$<$\,1.9, 1.5\,$\leq$\,a\,$<$\,3.0, 1.9\,$\leq$\,a\,$\leq$\,4.0, and 3.0\,$\leq$\,a\,$\leq$\,5.2 Gyr. The age groups are a reasonable compromise between the distributions in age and distance, number of objects, and covered age range. The six oldest open clusters are not consider in the analysis, because they will significantly enlarge the covered age range of the oldest age group, and are too small in number to represent an additional age group. Furthermore, we exclude the two most distant objects Berkeley~29 and Saurer~1 (see Fig. \ref{fig:merged}), because of their special Galactic location about 5\,kpc more distant than other clusters in the sample. A short discussion of these objects is given in Sect. \ref{sect:special-objects}.

We derived the metallicity gradients of the age groups using a robust regression model \citep{1981rost.book.....H}, already applied in Sect. \ref{sect:rmd}, and a bootstrap method with 10\,000 resampling iterations and linear least-squares regression. We note that the bootstrap distributions are almost Gaussian.
     
\begin{figure}
\centering
\resizebox{\hsize}{!}{\includegraphics{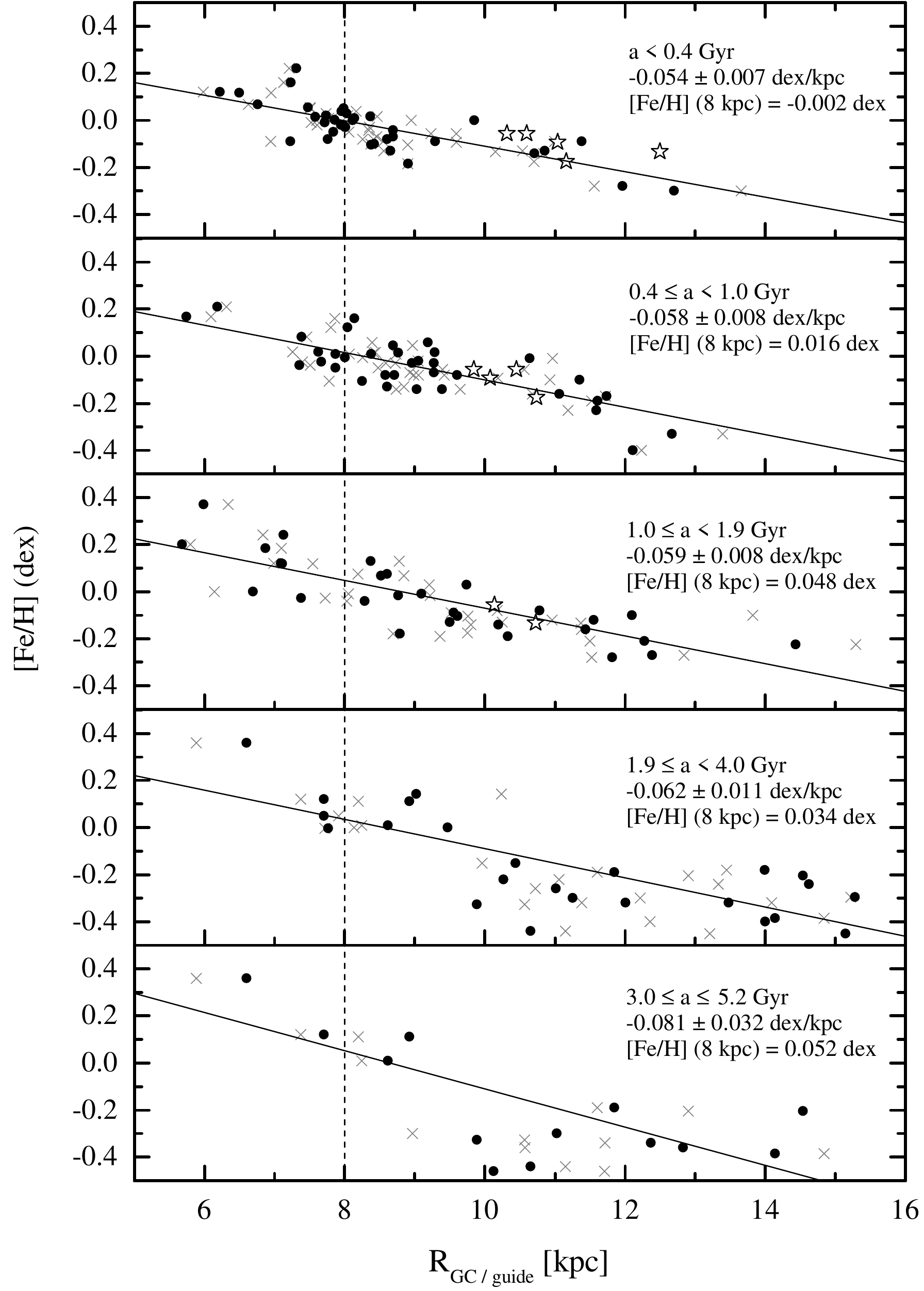}}
\caption{The metallicity distribution of five age groups. Overplotted are the derived gradients for \rgc\ using the robust fitting model. We also indicate their results and the metallicity levels at the solar circle. The photometric results are included as asterisks to show the good agreement with spectroscopic data. The \rguide\ positions are always given as crosses.}
\label{fig:agegroups}
\end{figure}

The results using \rgc\ as distance are listed in the upper panel of Table \ref{tab:gradients} and some age groups are shown in Fig. \ref{fig:agegroups}. All but the oldest group include about two to three dozens of objects and they cover distances out to about 13\,kpc at least. 

% ---------------------------
% ---------------------------

% ---------------------------
\begin{table*}
\caption{Metallicity gradients of the age groups  } 
\label{tab:gradients} 
\begin{center} 
\begin{tabular}{lccccccc} 
\hline
Age (Gyr) & N   & ZP  & \dexkpc & $\sigma$   & ZP  & \dexkpc & $\sigma$\\ 
\hline 
0.25 $\pm$ 0.11 & 41 & 0.43 $\pm$ 0.06 & $-0.054 \pm 0.007$ & 0.05 & 0.43 $\pm$ 0.06 & $-0.054 \pm 0.007$ & 0.06 \\

0.45 $\pm$ 0.13 & 29 & 0.46 $\pm$ 0.09 & $-0.056 \pm 0.009$ & 0.07 & 0.47 $\pm$ 0.08 & $-0.057 \pm 0.008$  & 0.07\\

0.74 $\pm$ 0.18 & 39 & 0.48 $\pm$ 0.07 & $-0.058 \pm 0.008$ & 0.07 & 0.49 $\pm$ 0.08 & $-0.059 \pm 0.008$ & 0.07 \\

0.98 $\pm$ 0.21 & 43 & 0.45 $\pm$ 0.07 & $-0.054 \pm 0.008$ & 0.09 & 0.50 $\pm$ 0.08 & $-0.059 \pm 0.009$ & 0.09 \\

1.32 $\pm$ 0.29 & 31 & 0.52 $\pm$ 0.07 & $-0.059 \pm 0.008$ & 0.09 & 0.54 $\pm$ 0.09 & $-0.061 \pm 0.008$  & 0.09\\

2.00 $\pm$ 0.40 & 25 & 0.49 $\pm$ 0.09 & $-0.058 \pm 0.008$ & 0.09 & 0.49 $\pm$ 0.08 & $-0.058 \pm 0.007$  & 0.09\\

2.54 $\pm$ 0.69 & 24 & 0.53 $\pm$ 0.13 & $-0.062 \pm 0.011$ & 0.14 & 0.52 $\pm$ 0.13 & $-0.061 \pm 0.011$  & 0.14\\

3.98 $\pm$ 0.82 & 13 & 0.70 $\pm$ 0.35 & $-0.081 \pm 0.032$ & 0.18 & 0.63 $\pm$ 0.31 & $-0.076 \pm 0.028$ & 0.18 \\

\\
0.25 $\pm$ 0.11 & 41 & 0.44 $\pm$ 0.05 & $-0.057 \pm 0.006$ & 0.06 & 0.46 $\pm$ 0.06 & $-0.058 \pm 0.007$ & 0.06 \\
0.45 $\pm$ 0.13 & 29 & 0.44 $\pm$ 0.05 & $-0.056 \pm 0.006$ & 0.05 & 0.47 $\pm$ 0.06 & $-0.058 \pm 0.006$  & 0.05\\
0.74 $\pm$ 0.18 & 38 & 0.52 $\pm$ 0.07 & $-0.063 \pm 0.008$ & 0.07 & 0.52 $\pm$ 0.08 & $-0.063 \pm 0.008$ & 0.07\\     
0.98 $\pm$ 0.21 & 42 & 0.50 $\pm$ 0.08 & $-0.059 \pm 0.009$ & 0.09 & 0.51 $\pm$ 0.11 & $-0.059 \pm 0.012$ & 0.09 \\
1.32 $\pm$ 0.29 & 31 & 0.47 $\pm$ 0.09 & $-0.055 \pm 0.010$ & 0.10 & 0.48 $\pm$ 0.11 & $-0.055 \pm 0.011$ & 0.10 \\
2.00 $\pm$ 0.40 & 25 & 0.43 $\pm$ 0.10 & $-0.054 \pm 0.009$ & 0.10 & 0.44 $\pm$ 0.09 & $-0.055 \pm 0.009$  & 0.10\\
2.54 $\pm$ 0.69 & 24 & 0.57 $\pm$ 0.12 & $-0.067 \pm 0.011$ & 0.12 & 0.57 $\pm$ 0.11 & $-0.067 \pm 0.010$ & 0.12 \\
3.98 $\pm$ 0.82 & 13 & 0.67 $\pm$ 0.20 & $-0.084 \pm 0.019$ & 0.15 & 0.68 $\pm$ 0.25 & $-0.083 \pm 0.024$  & 0.15\\                           
\hline
\end{tabular}
\end{center}
\flushleft
\medskip
\textit{Notes}. The left-hand side represents the results based on robust regression, and the right-hand side the results using a bootstrap method. The upper panel provides the gradients based on \rgc\ and the lower one using \rguide. The median age of each group is given in the first column, also the number of objects (N) and the chemical dispersion in each age group ($\sigma$) is listed. We note that we were not able to derive \rguide\ for one object (Berkeley~104) that is included in age groups 3 and 4.
\end{table*}
% ---------------------------

The results of the two approaches agree very well and indicate within the errors an almost constant metallicity gradient. The oldest age group ($\sim 4$\,Gyr), however, is showing a steeper gradient, but the errors and the small number of objects involved does not allow a firm conclusion. This result agrees very well with \citet{2011A&A...530A.138C} and \citet{2017A&A...600A..70A} for field stars. These indicate a flattening of the gradient only for older ages, which are not yet sufficiently covered by open clusters. 
Apart from the analysis of the defined age groups, we also performed a robust multiple linear regression on 130 clusters younger than about 3\,Gyr by including also the age as independent variable (the errors of the last significant digits are given in parenthesis):

\begin{equation}
$\feh\ = 0.49(4) - 0.060(4)\,\rgc\ + 0.020(10)\,Gyr ($\sigma$ = 0.08\,dex).$
\end{equation}

\citet{2016A&A...585A.150N} list gradients for two analysed age groups that are considerable larger than ours. This is a result of a much narrower covered \rgc\  range where inaccurate data have a stronger influence on the derived gradient, the inclusion of some cluster metallicities that are actually based on non-members (see Sect. \ref{datacompilation}), but also based on differences in the adopted cluster age and distance. \citet{2016A&A...585A.150N} use mean literature results, which generally reduces the influence of inaccurate parameters. However, this approach is probably not suited for little studied objects. One example is NGC~2354, for which a mean age of about 0.2\,Gyr and a distance of 3.6\,kpc is listed. However, the adopted results by \citet{2020A&A...640A...1C} quote about 1.4\,Gyr and 1.4\,kpc based on \textit{Gaia} data. We inspected the \textit{Gaia} CMD and conclude that \citet{2020A&A...640A...1C} provide a reasonable parametrization for this object. This clearly shows the advantage of the availability of \textit{Gaia} photometric and astrometric data for open cluster research, in particular to identify true giant star members to better constrain the age. 

The shallower gradients that we derive for the youngest age groups agree very well with results based on Cepheid stars \citep{2014A&A...566A..37G}, who derived $-$0.051 to $-$0.06 \dexkpc\ using different samples. \citet{2017A&A...600A..70A}, on the other hand, used red giant stars to derive $-$0.058 \dexkpc\ (their bias-corrected result) for the youngest group ($<$\,1\,Gyr). 

A step-like discontinuity of the metallicity gradient as proposed by \citet{2011MNRAS.417..698L} can be generally ruled out for the youngest age groups. For the older groups, however, one might identify such a jump of the metallicity level. Though  this appearance is very probably just caused by the lower number of objects and the stronger influence of migration effects in these groups, which cover a much larger age range. Resonances have certainly an impact on the clusters' distribution as shown by \citet{2021FrASS...8...62M}, but these effects must be observed more clearer among the youngest objects that have not moved much from their birth places.

\citet{2020A&A...634A..34B} presented a new approach to derive abundances for young dwarf stars in open clusters. They noticed that the microturbulence parameter is overestimated if using a ``standard'' analysis, resulting in an underestimate of the metal content. The largest differences were found for clusters younger than 100\,Myr. We note that their results are not included in our sample, though all their objects, except NGC 2547, for which we derive an offset of about 0.1\,dex. Thus, the result for our youngest cluster population might be affected. However, our youngest group includes only eight objects that are younger than 100\,Myr and 12 objects in total with a metallicity that is based on dwarf stars. 
Even the exclusion of all objects younger than 150\,Myr results in a gradient of $-$0.054 \dexkpc\ and a metallicity level at the solar circle of $-$0.012\,dex for the youngest age group, in agreement to the results listed in Table\,\ref{tab:gradients} and shown in Fig. \ref{fig:agegroups}. The median age of the remaining  objects is only slightly increased to 280\,Myr. 

Recently, \citet{2021MNRAS.503.3279S} used combined GALAH \citep{2021MNRAS.506..150B} and APOGEE data for a study of the metallicity gradients. They used 134 open clusters, a sample of similar size to ours. However, except one object all of them are closer than \rgc\,$\sim$\,13\,kpc (adopting $\rm R_{\odot}$\,$\sim$\,8.2\,kpc). They conclude that the guiding radius \rguide\ is a better distance indicator for metallicity gradient analyses and reduces effects by blurring. We therefore repeated the previous calculations using \rguide\ instead of \rgc. Although the use of \rguide\ provides a smoother metallicity distribution for some ``outliers'', some others pop up as new deviating objects. The results in the lower panel of Table\,\ref{tab:gradients} show that the gradients do not differ significantly, these agree very well within the errors. The chemical dispersion does not significantly improve as well, except in the very oldest group, in some age groups the use of \rguide\ even results in marginal larger dispersions. Furthermore, a multiple linear regression provides almost the same result as for \rgc:

\begin{equation}
$\feh\ = 0.49(3) - 0.061(4)\,\rguide\ + 0.018(9)\,Gyr ($\sigma$ = 0.07\,dex).$
\end{equation}

From fig.\,7 by \citet{2021MNRAS.503.3279S}, we infer that the metallicity gradient of their second age group (1 -- 2\,Gyr) does not change, while the others show shallower gradient by 0.02 -- 0.06\,\dexkpc\ using \rguide\ as distance indicator. In contrast, the overall gradient remain almost the same using \rgc\ ($-$0.076 \dexkpc) or \rguide\ ($-$0.073 dex/kpc). This generally can be explained if one group shows a better distance range covering or contributes most to the complete sample. Furthermore, for the youngest groups, one would expect the least influence by blurring \citep[see e.g.][]{2021A&A...647A..19T}, which is also supported by the results in Table\,\ref{tab:gradients}. The differences between our study and \citet{2021MNRAS.503.3279S} might be related with their broad age groups or insufficient distance coverage in the individual age groups, but the main difference is the adopted distance scale. \citet{2021MNRAS.503.3279S} used \textit{Gaia} parallax data to infer the cluster distances. These are certainly useful for closer objects, but the distances of objects more distant than $\sim$\,3\,kpc are continuously underestimated -- Berkeley~20 and Berkeley~29 even by more than 4\,kpc compared to our scale. This certainly produces steeper metallicity gradients using \rgc, but it is out of scope to evaluate the influence of the underestimated distances on \rguide. Thus, a comparison of metallicity gradients from different studies is complicated by the use of different distance scales, but also by the covered range of galactocentric distances and the applied method to derive the gradients.

\subsection{Radial migration} 

Radial migration is already a well-known effect, but it is still not well understood. Available chemodynamical models significantly differ in their approach:  e.g. \citet{2009MNRAS.396..203S} introduced a parametrization by hand, while \citet{2013A&A...558A...9M} used a cosmological simulation. However, radial migration appears important even for older kinematic hot populations.
 
An increase of the metallicity level at the solar circle with age can be noticed in several open cluster studies \citep[e.g.][]{2016A&A...591A..37J,2016A&A...585A.150N,2020AJ....159..199D,2021MNRAS.503.3279S}. \citet{2016A&A...585A.150N} conclude that this metallicity shift might be related with radial mixing. This behaviour seems independent on the used data and is also clearly present in our study. The derived metallicity gradients indicate a constant increase of the metallicity level at the solar circle for clusters up to about 2\,Gyr (see Fig.\,\ref{fig:agegroups}). This is supported by the results of the multiple linear regressions in Sect.\,\ref{sect:amr}; thus, it is not artificially caused by a possible inappropriate grouping. The simulation by \citet{2017A&A...600A..70A} shows that radial mixing might produce a metallicity shift of about +0.1\,dex after $\gtrsim$\,2\,Gyr by bringing metal-rich objects from inner regions to the solar circle. This metallicity increase is obviously about twice as large than supported by our data, though one has to keep in mind that the metallicity gradient and metallicity level of the ISM has changed in the past. Fig. 5 by \citet{2018MNRAS.481.1645M} indicates that the metallicity level at the solar circle was about 0.1\,dex lower 3\,Gyr ago and the gradient about 0.01\,\dexkpc\ steeper 4\,Gyr ago. 

We adopt the model by \citet{2018MNRAS.481.1645M} and used our youngest age group as a basis for the current ISM to estimate the birth radii (\rbirth). For the youngest age group one can assume that these clusters have not moved much since birth, but erroneous metallicities might still alter the derived gradients. We therefore checked the results by excluding few objects that show the lowest weight ($<$\,0.7) in the robust fitting procedure. This removed five and four clusters from the \rgc/\rguide\ samples, respectively. The clearly lowest weight ($<$\,0.4) was assigned to NGC~6087 in both samples. The cleaned data sets provide in both cases a gradient of $-$0.055(5)\,\dexkpc\ in good agreement with the results in Table\,\ref{tab:gradients}. We note that for older age groups, which are probably already affected by radial migration, such a procedure could be problematic as it also might reduce the scatter caused by radial migration.

We finally derive the migration distances \rgc\ $-$ \rbirth\ and \rguide\ $-$ \rbirth\ in two ways, first using the gradients of the individual age groups (adopting the solar circle as reference point), but also for all individual objects. However, we note that it is impossible to derive the true birthplaces as one has to assume that the ISM gradient has zero scatter.      
 
\begin{figure}
\centering
\resizebox{\hsize}{!}{\includegraphics{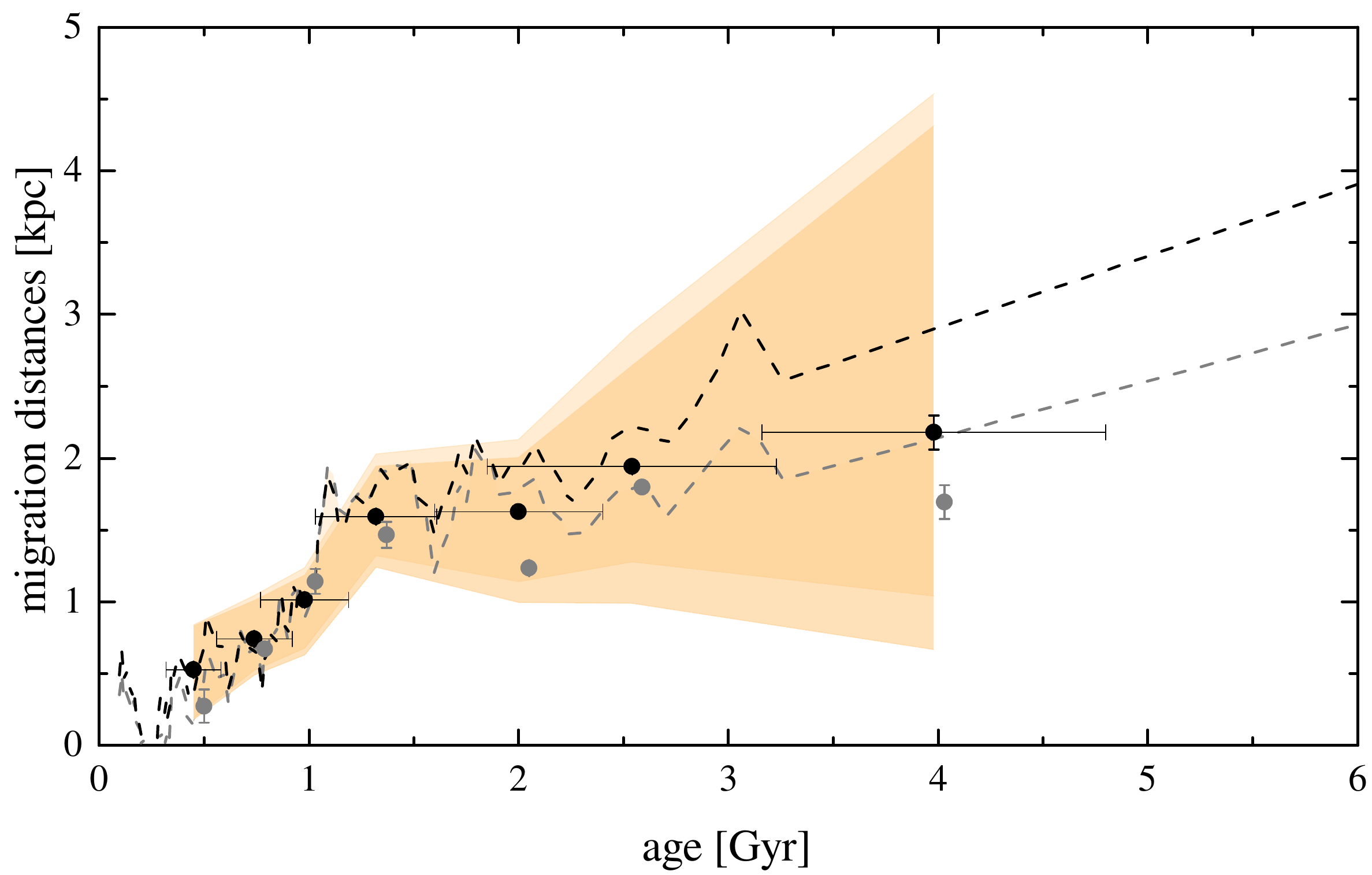}}
\caption{The required mean movement of the clusters by migration as a function of age. We provide the mean value derived from the two gradient sets (robust fitting and bootstrap method). The inner area shows the error range by considering the uncertainty of the regression analysis, the outer area includes also the error in age. The youngest age group was omitted as it deals as the reference for the present ISM. The dashed lines represent the running average on the individual migration distances. The black symbols and line represent the migration distance \rgc\ $-$ \rbirth, the grey ones \rguide $-$ \rbirth.}
\label{fig:radialmigration}
\end{figure}

Figure \ref{fig:radialmigration} shows the required mean movement by migration from the inner disc as a function of age by considering the present-day metallicity levels and gradients of each mono-age population and the changes of the ISM metallicity level and gradient in the past. We derived the migration distances using the two gradient sets (obtained by robust fitting and the bootstrap method), which agree very well. Furthermore, we present a running average of the migration distances derived for the individual objects. Here we use a grouping of 15 open clusters up to an age of 4\,Gyr, the remaining 11 oldest objects represent a separate group. 
We present the results using the two distance indicators \rgc\ and \rguide. Whereas the first provides an estimate of the total radial mixing by churning and blurring, \rguide\ gives an indication for churning only. 

% ---------------------------
% ---------------------------

For clusters younger than about 1.5\,Gyr we notice a mean migration rate of 1\,\kpcgyr. This agrees with a radial migration model \citep{2018MNRAS.475.4450Q} based on shearing sheet \textit{N}-body simulations, which is suitable for younger objects up to about 1\,Gyr (or about five orbital periods). They quote a maximum migration rate of 1\,\kpcgyr\ for a pitch angle of 24$^{\circ}$, which shows a good agreement with our results for the youngest clusters. However, they note that this migration rate is probably underestimated, because they derived higher rates for some open clusters. Our results probably include some underestimate as well. Recently, \citet{2021A&A...647A..39S} identified a star formation burst about 0.5\,Gyr ago. Assuming that this burst increased the overall metallicity level of subsequent generations, thus in particular of our youngest age group that was used as a reference, our calculations start from a somewhat too high metallicity level.

For older clusters different conclusions might be drawn from the results of the two analysis methods. The results of the gradient analysis would indicate that clusters almost do not migrate after 1.5\,Gyr and that the little additional migration distances are just caused by blurring effects. This conclusion, however, rest on the oldest age group at 4\,Gyr, which shows quite large errors for the gradient fits. A marginally flatter gradient would bring the migration distance already in line with the results obtained from the individual objects. These data on the other hand suggest that for older (2-6\,Gyr), dynamically hotter, populations the total mean migration rate drops to $\sim$\,0.5\,\kpcgyr\ with a contribution by churning of 0.3\,\kpcgyr. Furthermore, at an age of 6\,Gyr, blurring already adds about 1\,kpc to the total migration distance. This is compatible with a model by \citet{2020ApJ...896...15F}, which suggests a radial epicycle spread of about 1.3\,kpc after 6\,Gyr. There are some limitations in these findings as well, because first it does not account for possible differences with the galactic location. Actually, the mean galactocentric distance increases constantly with age from 8 to 12\,kpc. Furthermore, the scatter of the migration distances is of the same order as the mean values. 

\begin{figure}
\centering
\resizebox{\hsize}{!}{\includegraphics{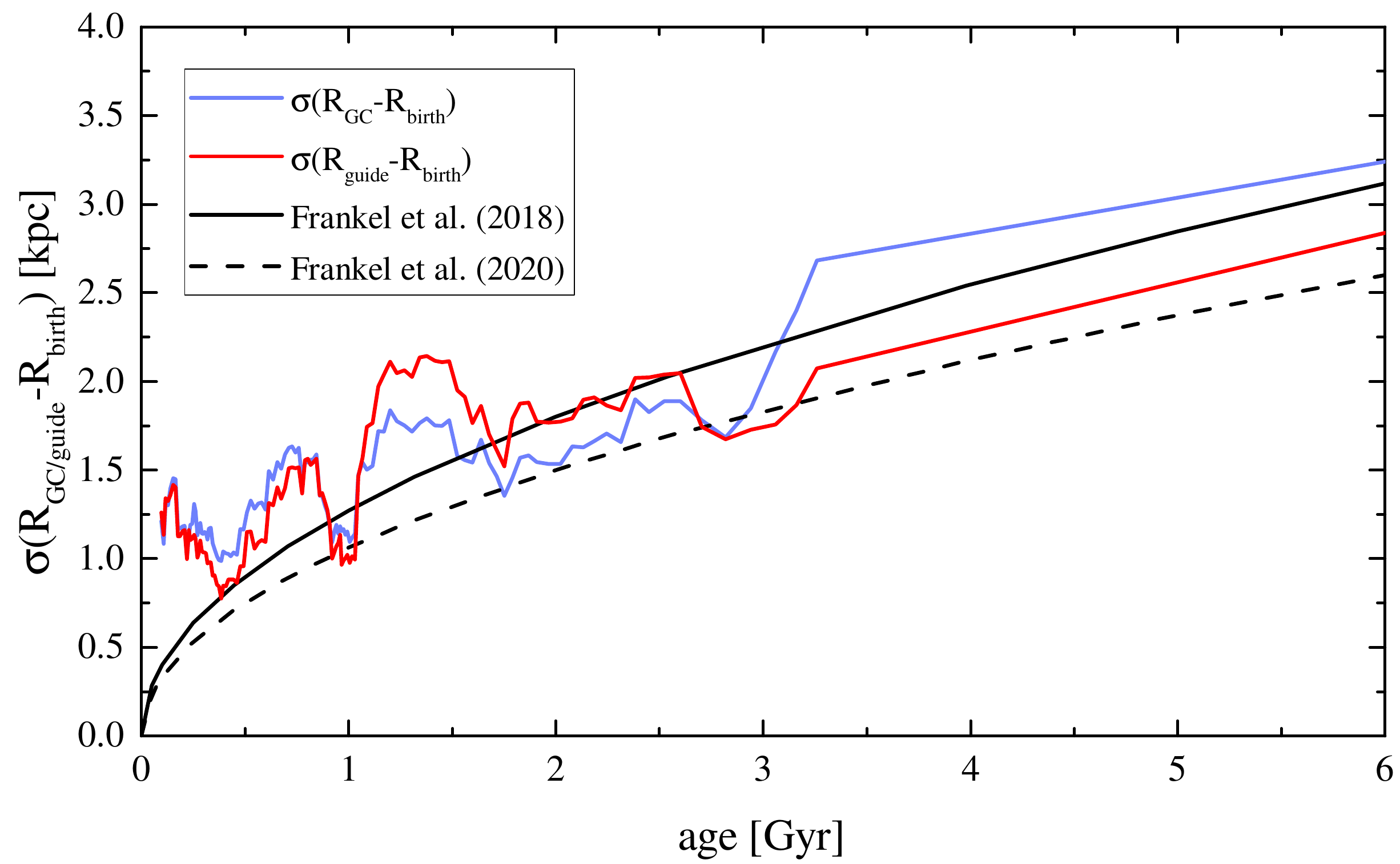}}
\caption{The evolution of the scatter, derived for \rgc\ and \rguide, compared with the radial migration efficiency models by \citet{2018ApJ...865...96F,2020ApJ...896...15F}.}
\label{fig:scatter}
\end{figure}

However, the evolution of the scatter provides a useful information as well. Radial migration efficiency models were fitted to red clump star data by \citet{2018ApJ...865...96F,2020ApJ...896...15F} to retrace their chemical dispersion with age. Their earlier model is purely spatial and does not include the dynamics \citep[see][]{2020ApJ...896...15F}. The difference in the models thus represents the influence by blurring. Figure\,\ref{fig:scatter} shows the evolution of the scatter in our data and compares it with the two models above. These generally follow the trend in our data; also, the increase of the scatter by blurring is well reproduced at least for an age $>$\,3\,Gyr. This agrees with the findings in Table\,\ref{tab:gradients}, where the two oldest groups show a noticeable reduction of the scatter between \rgc\ and \rguide.

% ---------------------------
% ---------------------------

A comparable approach to ours was adopted by \citet{2020MNRAS.495.2673C}. They used the sample by \citet{2016A&A...585A.150N} including also their photometric data compilation, which contributes about 40 per cent to the total sample. They note that blurring causes significant radial excursions for only 10 per cent of the clusters, which are mainly located in outer Galactic regions. Furthermore, they found that among the migrating objects younger clusters (<\,1\,Gyr) tend to migrate inwards while older ones migrate outwards. The photometric sample by \citet{2016A&A...585A.150N} without any spectroscopic data covers mostly younger objects ($\sim$\,70 per cent <0.5\,Gyr) close to the solar circle. For the majority of these objects also only a single photometric metallicity estimate from various methods is available and their reliability was questioned by \citet{2016A&A...585A.150N}. We therefore conclude that in particular the conclusion by \citet{2020MNRAS.495.2673C} for their young group is probably strongly biased by photometric data. 

\citet{2021ApJ...919...52Z} on the other hand compiled only spectroscopic data and conclude that 46 per cent of open clusters migrated more than 1\,kpc and that older clusters generally migrate outward. In agreement with our finding, they derived a migration rate of about 1\,kpc/Gyr for the youngest clusters. Furthermore, using their criterion for migrators, we derive an almost identical value of 51 per cent. 

Considering the uncertainty introduced by the ISM model in our calculations and the limitation by using a 0.25-Gyr-old population as a proxy for the current ISM, we conclude that the observed increase of the metallicity levels with age can be quite well explained by radial migration itself. 
There is actually no need to introduce other explanations as main solution for a lower metallicity level of young clusters, such as systematics in the spectroscopic analysis due to stellar activity, for example. Using for example the models by \citet{2018MNRAS.481.1645M} and \citet{2018MNRAS.475.4450Q}, one must observe some increase of the metallicity level, because the resulting shift in the metallicity is about a factor of 2 larger than the decreasing one with age of the ISM model. In other words, a decrease of the metallicity levels with age might be only noticed if the radial migration efficiency is lower than about 0.5\,kpc/Gyr.

\citet{2017A&A...600A..70A} proposed a mechanism including radial mixing and a more prone tidal disruption of inward-migrating objects
to explain observed discrepancies between the gradient evolution of clusters and field stars. In our sample we notice only 12 inward-migrating
objects (most of them younger than 1\,Gyr) using (\rguide\,$-$\,\rbirth)\,$<$\,$-1$\,kpc, and 49 per cent probable non-migrators using $|$\rguide\,$-$\,\rbirth$|$\,$<$\,1\,kpc. The low number
of inward-migrating objects might support their proposal. However, our results for the gradients (Table \ref{tab:gradients}) are quite constant over time and
reasonably agree with results for field stars \citep{2011A&A...530A.138C,2014A&A...566A..37G,2017A&A...600A..70A}. An exception is the oldest age group (4\,Gyr), for which we still find a somewhat steeper gradient, but the low number of covered objects and the large error of the gradient does still not allow
a firm conclusion for this age range -- see e.g. also the discussion by \citet{2017A&A...600A..70A} for their little covered oldest age group ($>$10\,Gyr).
On the other hand, the evolution of the scatter of the migration distances (Fig.\,\ref{fig:scatter}) agrees very well with the results by \citet{2018ApJ...865...96F,2020ApJ...896...15F}, suggesting that there might be no peculiarity of the open cluster population. Future studies are clearly needed to improve the coverage of the oldest open clusters.

\subsection{Objects of special interest}
\label{sect:special-objects}

The current Galactic position of the most distant open clusters in our sample (Berkeley~29 and Saurer~1) does absolutely not fit to the radial metallicity distribution of the other clusters with a comparable age. Their orbits span a wide range in distance (see Fig. \ref{fig:merged}) and only their pericentre radii (or very close to them) would agree with the other clusters' metallicity distribution. Furthermore, their current position above the Galactic plane ($>$\,1.5\,kpc) is close to their orbital boundary value $\rm z_{max}$. \citet{2009MNRAS.397L.106C} speculated about an extragalactic origin of these objects, \citet{2016A&A...588A.120C}, on the other hand, note that the mean abundances place them at the limit between the $\alpha$-rich thin disc and the $\alpha$-poor thick disc without strong evidence of an extragalactic origin, but their orbits might be perturbed by accretion events or minor mergers in the past. In case of the last, it will be challenging to conclude about their actual birth radii.  

In the following, we discuss deviating objects using the weight of the applied robust fitting method as indicator.
Figure \ref{fig:merged} shows three objects with an apparent metal deficiency compared to other clusters at the same distance (\rgc\ $\sim$ 10 kpc): NGC~2243, Melotte~66, and Trumpler~5 (age $\sim$\,4-5\,Gyr). These were identified as deviating objects in Sect. \ref{sect:rmd}, and two of them (NGC~2243 and Trumpler~5) also in the analysis of the age dependence of the radial \feh\ gradient (Sect. \ref{sect:amr}). For all three clusters, their apocentre radii are in better agreement with the radial metallicity distribution of other objects in their age group. For two about 0.7-Gyr-old objects (NGC~2632 and FSR~716), an eccentric orbit already might explain their deviation, but for NGC~6087 and NGC~6583 there might be some issue concerning age, distance, metallicity scale, or they show a quite different radial migration efficiency. 

A check of the \textit{Gaia} CMDs does not show noticeable problems with their adopted distance and age. The about 100-Myr-old cluster NGC~6087 is only included in the very youngest age group, suggesting that an unsuitable age group is very probably not the reason for the deviation. However, the adopted LQ metallicity of 0.22\,dex is based on a single (Cepheid) star measured by \citet{1994ApJS...91..309L}. An additional literature search noticed a newer result for the star by this author \citep{2018AJ....156..171L} with a lower metallicity: 0.11\,dex.  Furthermore, a photometric estimate of the cluster using the DG method gives 0.04\,dex \citep{2016A&A...585A.150N}.  

The metal-rich open cluster NGC~6583, on the other hand, deviates in two overlapping age groups. \citet{2018MNRAS.475.4450Q} conclude for this about 1.2-Gyr-old inner disc open cluster, currently located at \rgc\ $\sim$ \,6\,kpc, that either the Milky Way bar ejected the object from the inner Galaxy or a radial migration of 2-3\,kpc/Gyr is required to explain the current position. This estimate is about a factor of 2--3 larger than the mean radial migration we found for objects of comparable age. We estimate for this cluster even a much larger migration distance of about 5\,kpc.

Another metal-rich inner disc cluster, mentioned as a special case in several previous studies, is NGC~6791. It is not included in our investigation of the age dependence of the radial \feh\ gradient because of the little number of very old objects in the sample. With an age of about 8.5\,Gyr, it even belongs to the very oldest open clusters known in the Galaxy \citep[see e.g. the catalogue by ][]{2020A&A...640A...1C}. This cluster shows the most eccentric orbit in our sample with a pericentre distance of \rgc\ $\sim$ \,4.6\,kpc (see Figure \ref{fig:merged}). Furthermore, due to its age, also the largest migration distance by churning was estimated (almost 7\,kpc). Thus, suggesting a birthplace close to the Galactic Centre as for NGC~6583, if their metallicity is correct.

\section{Summary and concluding remarks}

This work presents a combined data set of spectroscopic metallicities for 136 open clusters in total. The cluster membership of a part of the sample, which was compiled in the pre-\textit{Gaia} era, was verified, resulting in the exclusion of some open cluster results that probably influenced previous conclusions based on these data.  The other important fundamental cluster parameters (distance and age) were evaluated in particular for the most distant objects in the sample to reduce the influence of inaccurate distances on the analysis of the metallicity gradients. 

Furthermore, we present a complete parametrization of 14 open cluster using a photometric method. These objects are located in a somewhat outer Galactic region (\rgc\ $\sim$ 10-12\,kpc) to improve the coverage of objects in this region. A good agreement with spectroscopic data was noticed, suggesting that the photometric approach provides a solid basis for future extensions. For example, we still notice a lack of a suitable coverage of metallicity among the oldest open clusters and in the outer disc. The total sample includes only about 20 objects older than 3\,Gyr and only a dozen clusters more distant than 13\,kpc from the Galactic Centre.   

The data were used to investigate the radial metallicity distribution, the age dependence of the radial \feh\ gradient using mono-age populations of eight consecutive and overlapping age groups using objects up to about 5\,Gyr, and to study the radial migration efficiency. Furthermore, we discuss some special objects of interest, which apparently do not follow the general trend in the metallicity distributions. 

We do not notice a significant change of the metallicity gradient out to \rgc\ $\sim$ 13\,kpc. A somewhat shallower gradient might be present beyond this radius, but the small number of covered objects in this Galactic region does not allow a firm conclusion yet. Though, based on the oldest clusters only one might conclude a flattening of the gradient already at about 10\,kpc in agreement with \citet{2021FrASS...8...62M}. However, the much smaller number of older objects requires the use of larger age ranges to define suitable covered mono-age populations, resulting in a larger scatter probably caused by radial migration effects. Thus, the metal content of more old clusters is needed to better pinpoint the trend and evolution of their metallicity gradients.    

The analysis of the age groups shows almost constant gradients, somewhat steeper gradients are noticed only for the oldest clusters, which however are little covered. Furthermore, we notice increasing metallicity levels with age, which we interpret as a result of a radial outward migration rate of 1\,kpc/Gyr for objects younger than $\sim$\,2\,Gyr. For older objects the results differ between the adopted methods (analysis of the gradients and using mean migration distances of the individual objects). The metallicity gradients indicate little additional migration, a conclusion very probably caused by the insufficiently covered oldest age group; the analysis of the individual objects, however, provides a total migration rate of about 0.5\,kpc/Gyr between 2 and 6\,Gyr, with churning somewhat stronger than blurring. The use of \rgc\ and \rguide\ as distance proxy indicates that the epicyclic excursions start to be recognisable after 2\,Gyr and contribute about 1\,kpc to the total migration distance after 6\,Gyr. A comparison of our migration distance estimates, but also their scatter, with radial migration models, shows good agreement.   

We still have to deal with an observational bias in the metallicity distribution. Most of the outer disc objects  (\rgc\ $\gtrsim$ 13\,kpc) with known metallicity are older than about 2\,Gyr. These are certainly easier to identify and to investigate because of their outlying position from the Galactic plane. Furthermore, the use of red giants are preferred in spectroscopic studies. Young distant objects are more obscured and are generally little known, the catalogue by \citet{2020A&A...640A...1C} only includes 24 objects younger than 1\,Gyr with a distance of \rgc\ $>$ 13\,kpc. In the solar neighbourhood, however, the age distribution shows that the younger ones represent the by far most frequent population \citep{2021A&A...645L...2A}. 

With increasing sample sizes one certainly also has to deal with an increasing number of objects that apparently do not fit into the overall picture. This might be simply related to inaccurate data, because most open cluster metallicities are based on few or even on a single star, or to specific peculiarities of the cluster. For some objects such as Berkeley~29 and Saurer~1, the most distant clusters in our sample, their orbit boundaries might explain their outstanding position in the metallicity distribution. For two clusters (NGC~6583 and NGC~6791), the known metallicity suggests a birthplace close to the Galactic Centre. For the much younger one (NGC~6583, 1.2\,Gyr), this already implies a migration rate of about 4\,Gyr/kpc. 
At least for one object (NGC~6087), the metallicity scale definitely requires further verification.

That \textit{Gaia} data have already significantly improved numerous topics in Galactic research is beyond doubt. Future data releases will also include homogeneous astrophysical parameters based on the full BP and RP spectrophotometry and even more precise using data from the higher resolution radial velocity spectrograph.
In combination with future individual spectroscopic or photometric studies, and ongoing or future large-scale spectroscopic surveys, the better sampling will also allow to switch from analyses using simple linear regressions to well-covered segmentations, to study regional properties of mono-age populations in even more detail.

\section*{Acknowledgements}
We thank the referee, Friedrich Anders, for valuable comments to improve this paper. MN would also like to thank Barbara, Axel, and Erik for their patience during preparation of this paper.
This paper is based upon observations carried out at the Observatorio Astron\'{o}mico Nacional on the Sierra San Pedro M\'{a}rtir (OAN-SPM), Baja California, M\'{e}xico. This paper has also made use of results from the European Space Agency (ESA) space mission Gaia, the data from which were processed by the Gaia Data Processing and Analysis Consortium (DPAC). Funding for the DPAC has been provided by national institutions, in particular the institutions participating in the Gaia Multilateral Agreement. The Gaia mission website is http://www.cosmos.esa.int/gaia. Furthermore, this research has made use of the WEBDA data base (https://webda.physics.muni.cz), operated at the Department of Theoretical Physics and Astrophysics of the Masaryk University.

%%%%%%%%%%%%%%%%%%%%%%%%%%%%%%%%%%%%%%%%%%%%%%%%%%
\section*{Data Availability}
The data underlying this paper are available at the CDS. These cover the complete Tables  \ref{tab:physparameter} and \ref{tab:orbitparameter}, and the derived differential grids.

%%%%%%%%%%%%%%%%%%%%%%%%%%%%%%%%%%%%%%%%%%%%%%%%%%



\bibliographystyle{mnras}
\bibliography{Netopil_metallicity} % if your bibtex file is called example.bib

%\begin{thebibliography}{99}
%\end{thebibliography}

%%%%%%%%%%%%%%%%%%%%%%%%%%%%%%%%%%%%%%%%%%%%%%%%%%

%%%%%%%%%%%%%%%%% APPENDICES %%%%%%%%%%%%%%%%%%%%%
\appendix
\section{}

% ---------------------------
\begin{table*}
\caption{Derived offsets of our photometric data.  } 
\label{tab:offsets} 
\begin{center} 
\begin{tabular}{l c c c c c c} 
\hline
Cluster & $\Delta V$   & $\Delta (V-R)$  & $\Delta (V-I)$   & $\Delta (V-R)$  & $\Delta (V-I)$   & $\Delta (B-V)$ \\ 
\hline 
Basel~4 & $-$0.02(0.04) & $-$0.03(0.03) & $-$0.05(0.04) & $-$0.09 & $-$0.06 & $-$0.02 \\
Berkeley~35 & +0.01(0.02) & +0.02(0.02) & $-$0.00(0.02) & +0.01 & +0.00 & $-$0.04 \\
Berkeley~60 & $-$0.02(0.04) & +0.03(0.04) & $-$0.04(0.04) & $-$0.05 & $-$0.04 & +0.00 \\
Berkeley~77 & $-$0.02(0.02) & $-$0.03(0.02) & $-$0.01(0.03) & $-$0.02 & $-$0.01 & $-$0.01 \\
Berkeley~104 & $-$0.00(0.02) & +0.02(0.02) & $-$0.02(0.02) & $-$0.03 & $-$0.03 & $-$0.04 \\
Haffner~4 & $-$0.00(0.03) & +0.02(0.04) & $-$0.01(0.03) & $-$0.01 & $-$0.01 & $-$0.09 \\
King~15 & +0.02(0.02) & +0.02(0.02) & $-$0.00(0.03) & $-$0.02 & $-$0.01 & $-$0.04 \\
King~23 & $-$0.01(0.02) & +0.00(0.02) & $-$0.03(0.03) & $-$0.01 & +0.00 & $-$0.02 \\
NGC~1857 & $-$0.00(0.02) & +0.02(0.02) & $-$0.03(0.03) & $-$0.02 & $-$0.03 & $-$0.05 \\
NGC~2186 & +0.00(0.03) & +0.00(0.03) & $-$0.04(0.04) & $-$0.03 & +0.00 & $-$0.04 \\
NGC~2236 & $-$0.00(0.02) & +0.02(0.02) & $-$0.02(0.02) & $-$0.02 &  $-$0.02 & $-$0.04 \\
NGC~2259 & +0.00(0.02) & +0.02(0.02) & $-$0.04(0.02) & $-$0.02 &  $-$0.04 & $-$0.06 \\
NGC~2304 & +0.02(0.02) & +0.01(0.02) & +0.02(0.02) & +0.01 &  +0.02 & $-$0.02 \\
NGC~2383 & $-$0.01(0.02) & $-$0.01(0.03) & $-$0.01(0.02) & $-$0.03 &  $-$0.01 & +0.00 \\
\hline 
\end{tabular}
\end{center}
\flushleft
\medskip
\textit{Notes}. The first three columns list the difference of transformed \textit{Gaia} photometry minus ours; the remaining columns give the derived offsets by obtaining a common temperature scale. All values are given in magnitudes, and the standard deviation is given in parentheses.
\end{table*}
% ---------------------------

% ---------------------------
\begin{figure*}
\centering
\resizebox{\hsize}{!}{\includegraphics{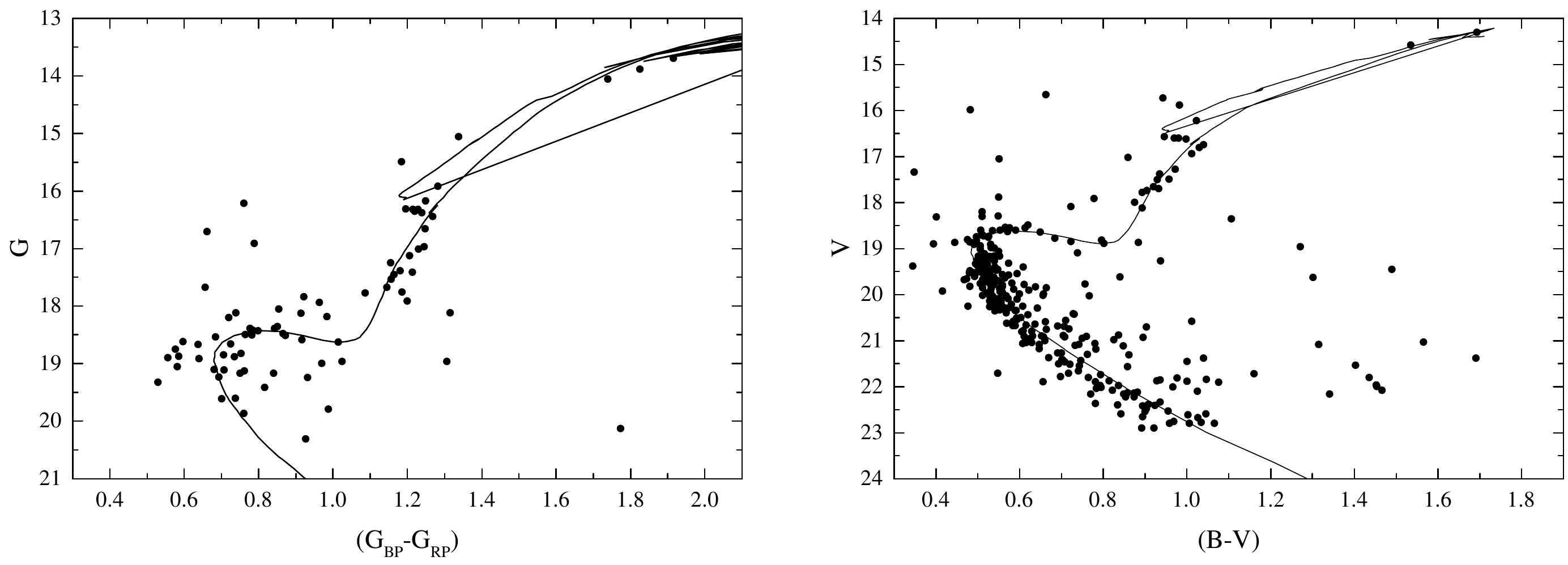}}
\caption{Result for the open cluster Berkeley~29. The left-hand panel shows the \textit{Gaia} CMD of proper motion members and the right-hand panel the $BV$ data by \citet{2004MNRAS.354..225T} restricted to the innermost cluster region. The overplotted isochrones correspond to the parameters listed in Table \ref{tab:dist-clusters}.}
\label{fig:be29}
\end{figure*}
% ---------------------------

% ---------------------------
\begin{figure*}
\centering
\resizebox{\hsize}{!}{\includegraphics{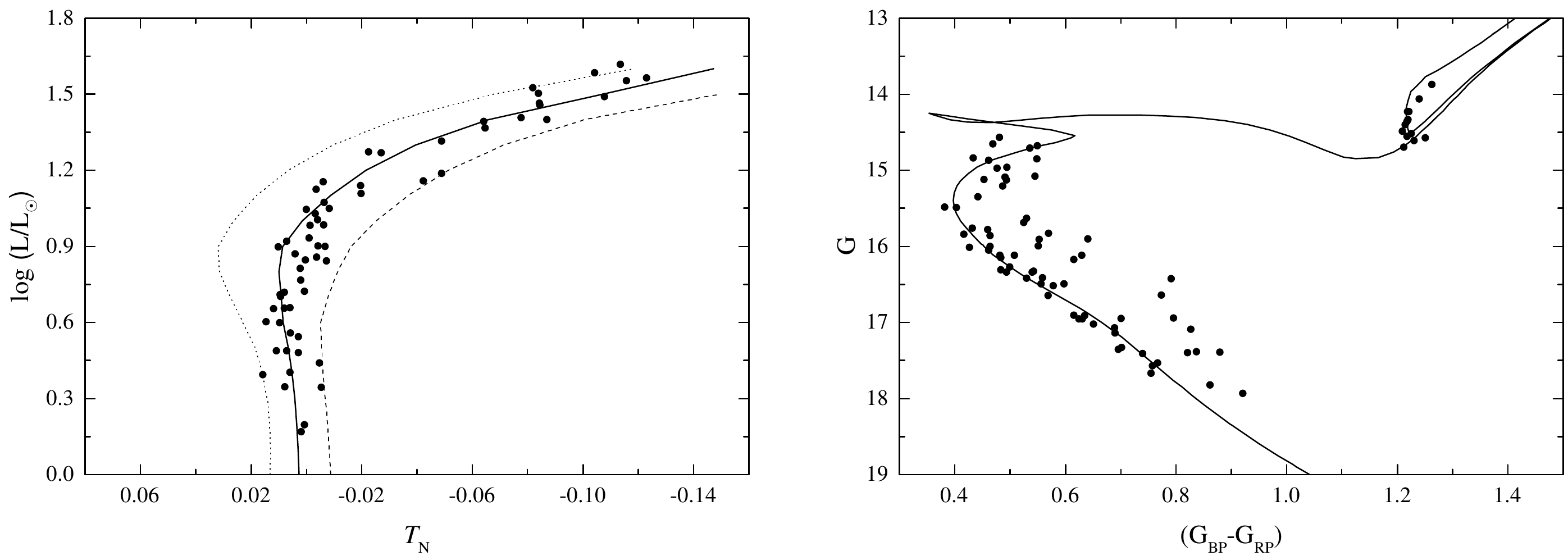}}
\caption{As Figure \ref{fig:basel4}, but for Berkeley~35.}
\label{fig:be35}
\end{figure*}
% ---------------------------

% ---------------------------
\begin{figure*}
\centering
\resizebox{\hsize}{!}{\includegraphics{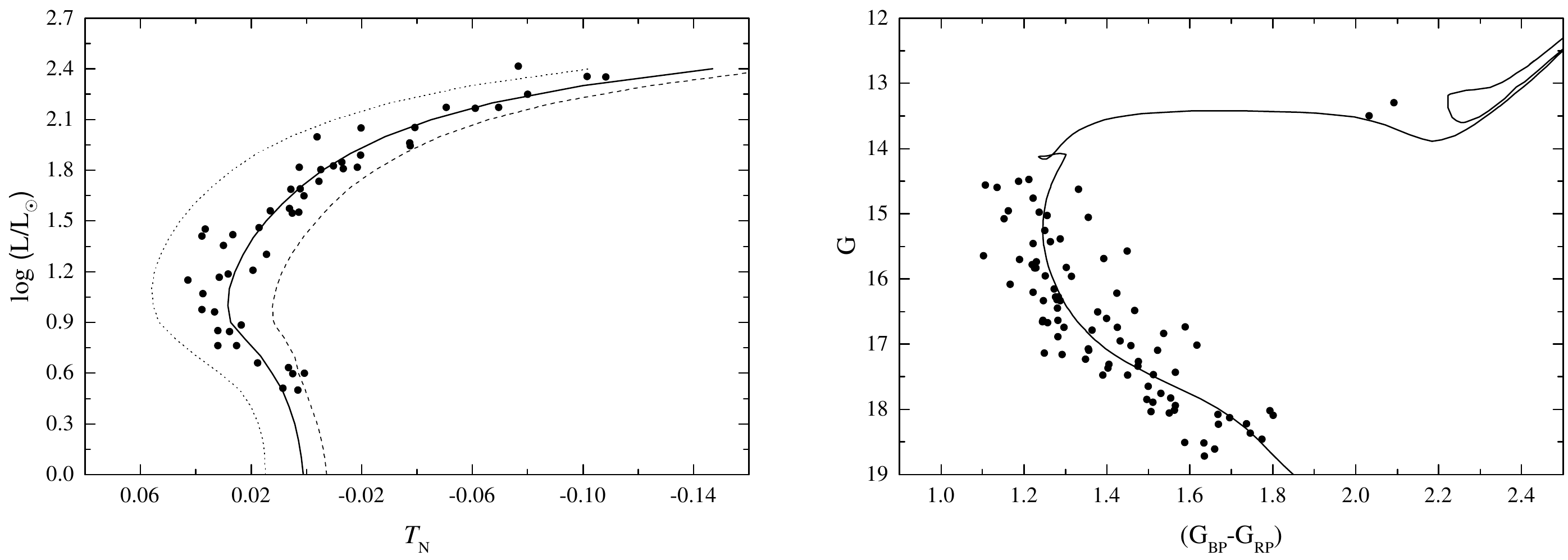}}
\caption{As Figure \ref{fig:basel4}, but for Berkeley~60.}
\label{fig:be60}
\end{figure*}
% ---------------------------

% ---------------------------
\begin{figure*}
\centering
\resizebox{\hsize}{!}{\includegraphics{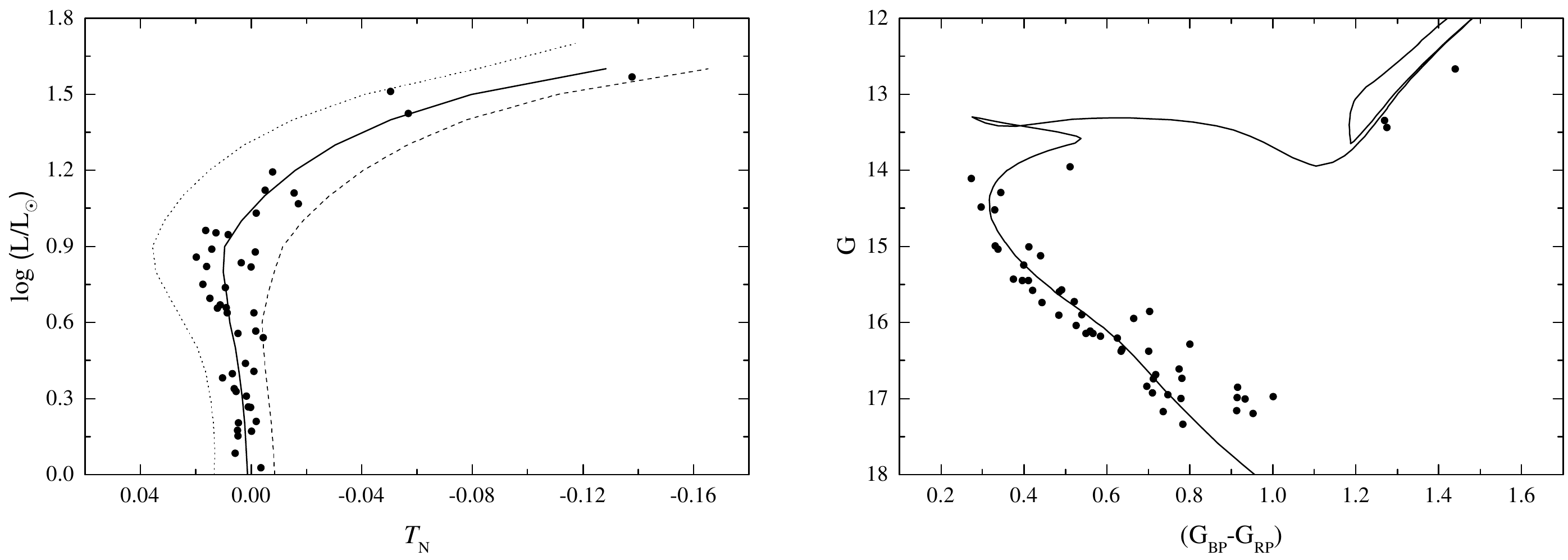}}
\caption{As Figure \ref{fig:basel4}, but for Berkeley~77.}
\end{figure*}
% ---------------------------

% ---------------------------
\begin{figure*}
\centering
\resizebox{\hsize}{!}{\includegraphics{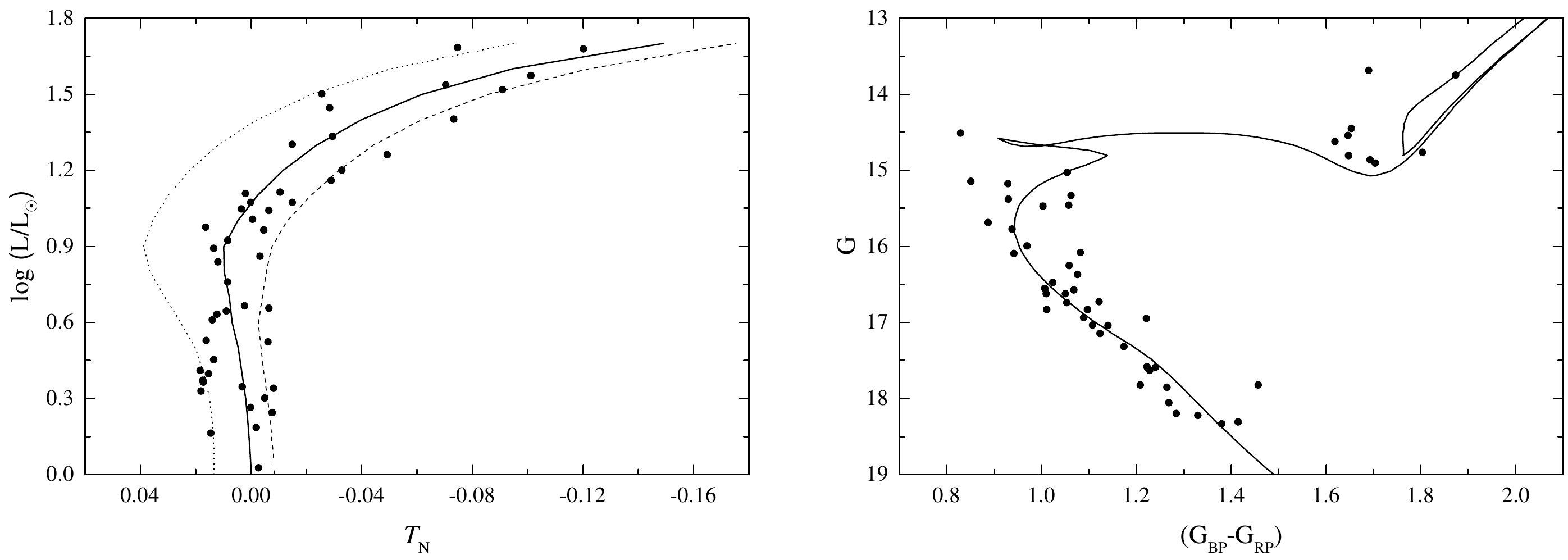}}
\caption{As Figure \ref{fig:basel4}, but for Berkeley~104.}
\end{figure*}
% ---------------------------

% ---------------------------
\begin{figure*}
\centering
\resizebox{\hsize}{!}{\includegraphics{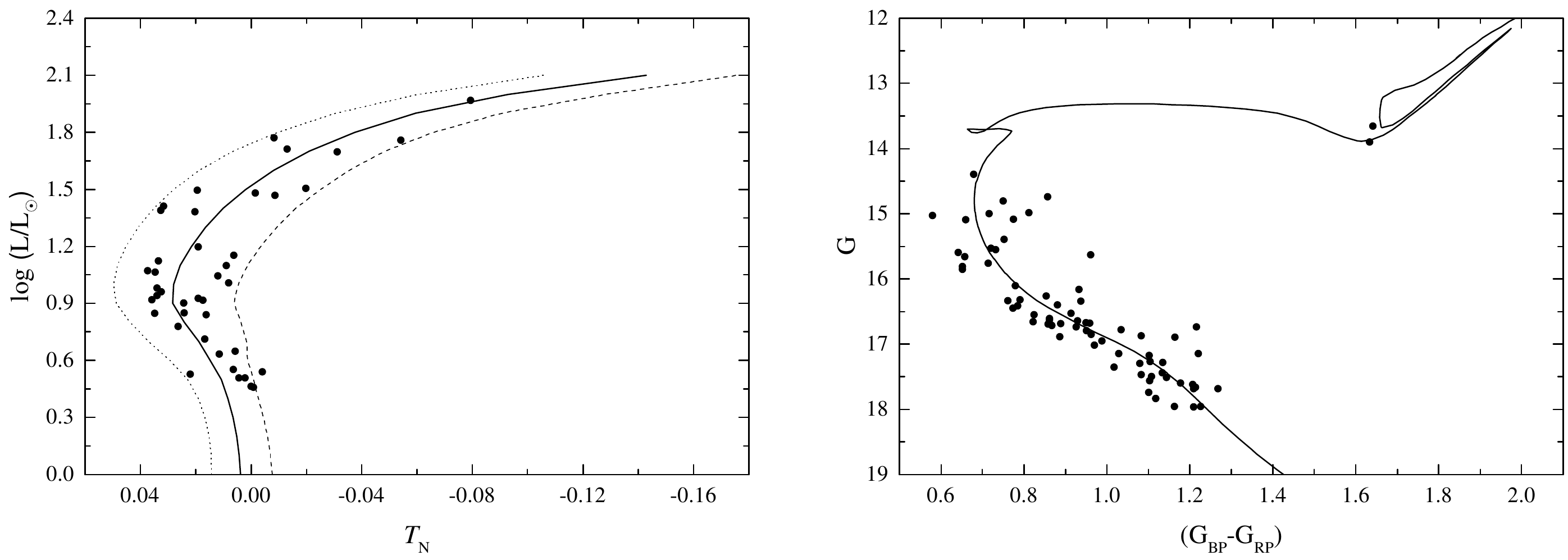}}
\caption{As Figure \ref{fig:basel4}, but for Haffner~4.}
\end{figure*}
% ---------------------------

% ---------------------------
\begin{figure*}
\centering
\resizebox{\hsize}{!}{\includegraphics{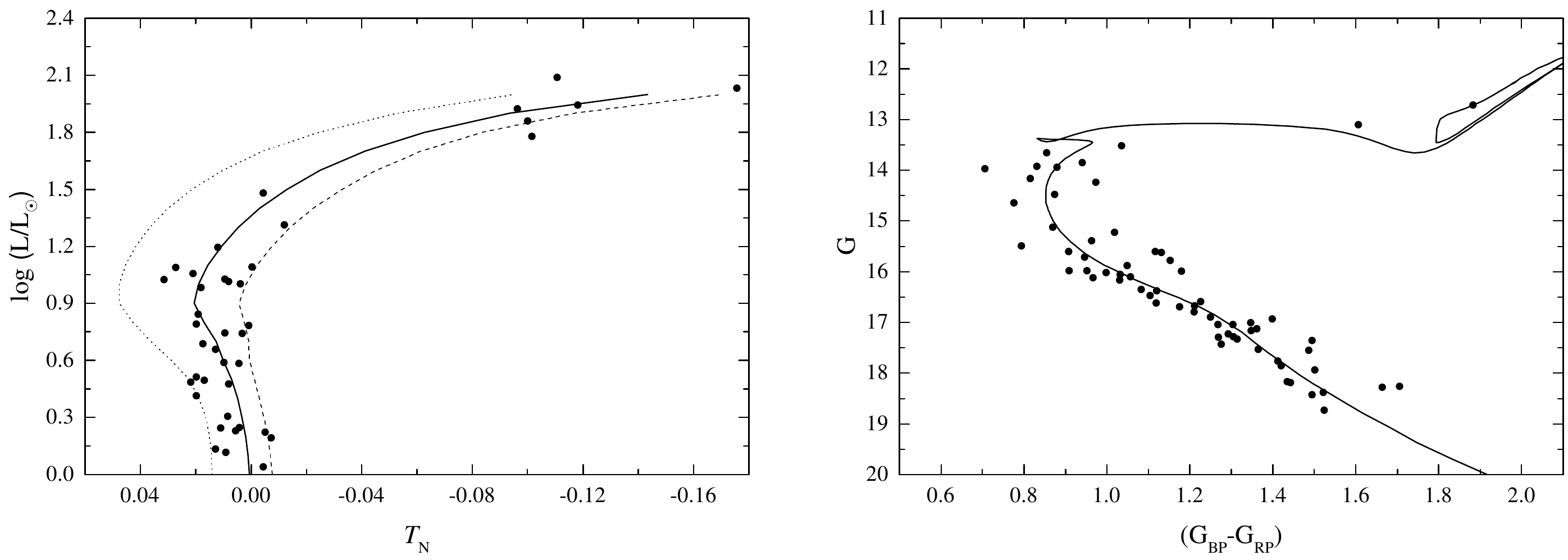}}
\caption{As Figure \ref{fig:basel4}, but for King~15.}
\end{figure*}
% ---------------------------

% ---------------------------
\begin{figure*}
\centering
\resizebox{\hsize}{!}{\includegraphics{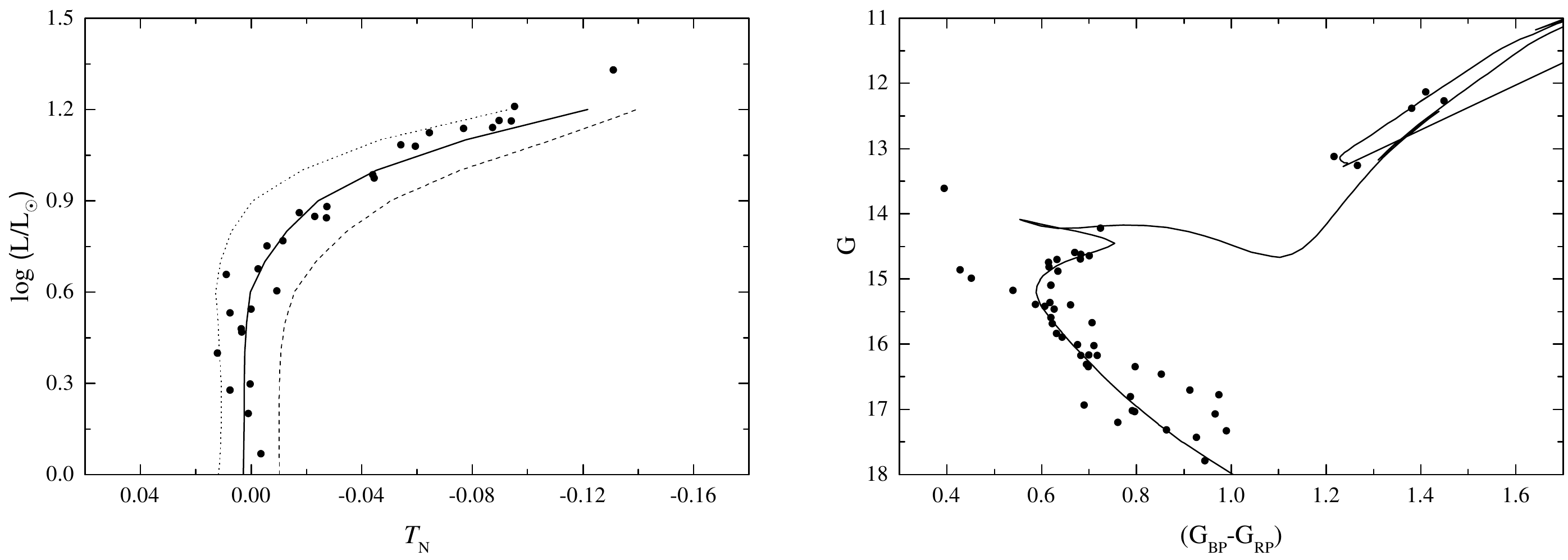}}
\caption{As Figure \ref{fig:basel4}, but for King~23.}
\end{figure*}
% ---------------------------

% ---------------------------
\begin{figure*}
\centering
\resizebox{\hsize}{!}{\includegraphics{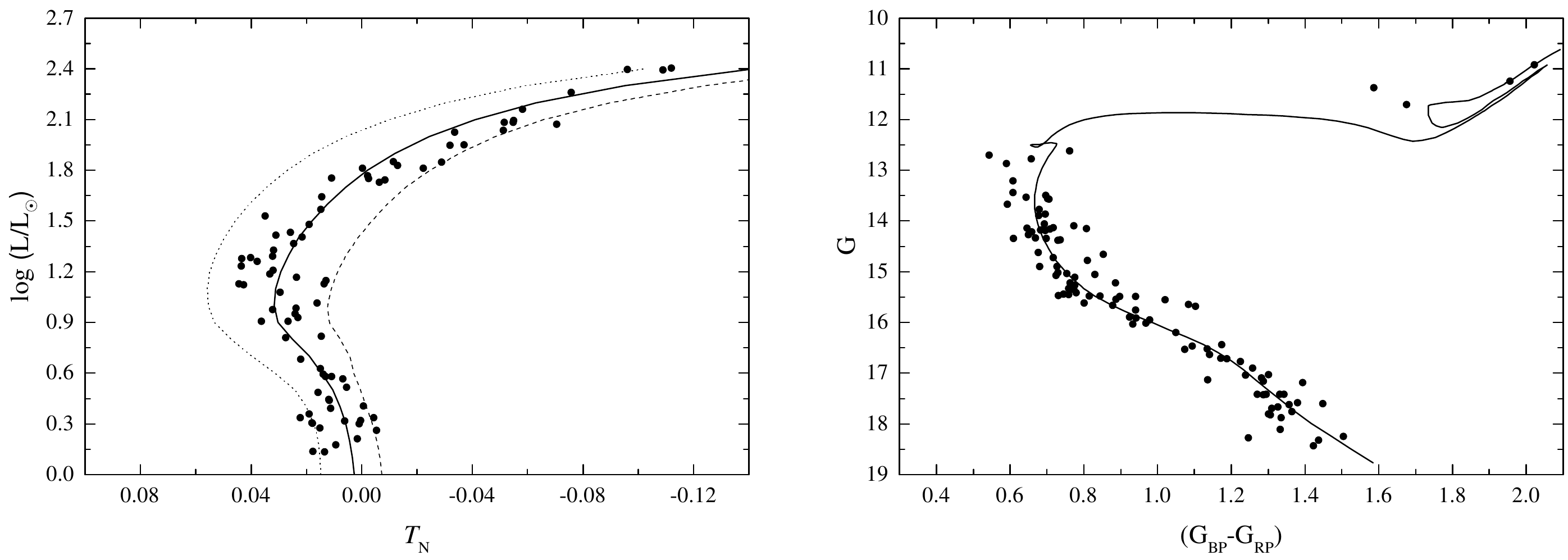}}
\caption{As Figure \ref{fig:basel4}, but for NGC~1857.}
\end{figure*}
% ---------------------------

% ---------------------------
\begin{figure*}
\centering
\resizebox{\hsize}{!}{\includegraphics{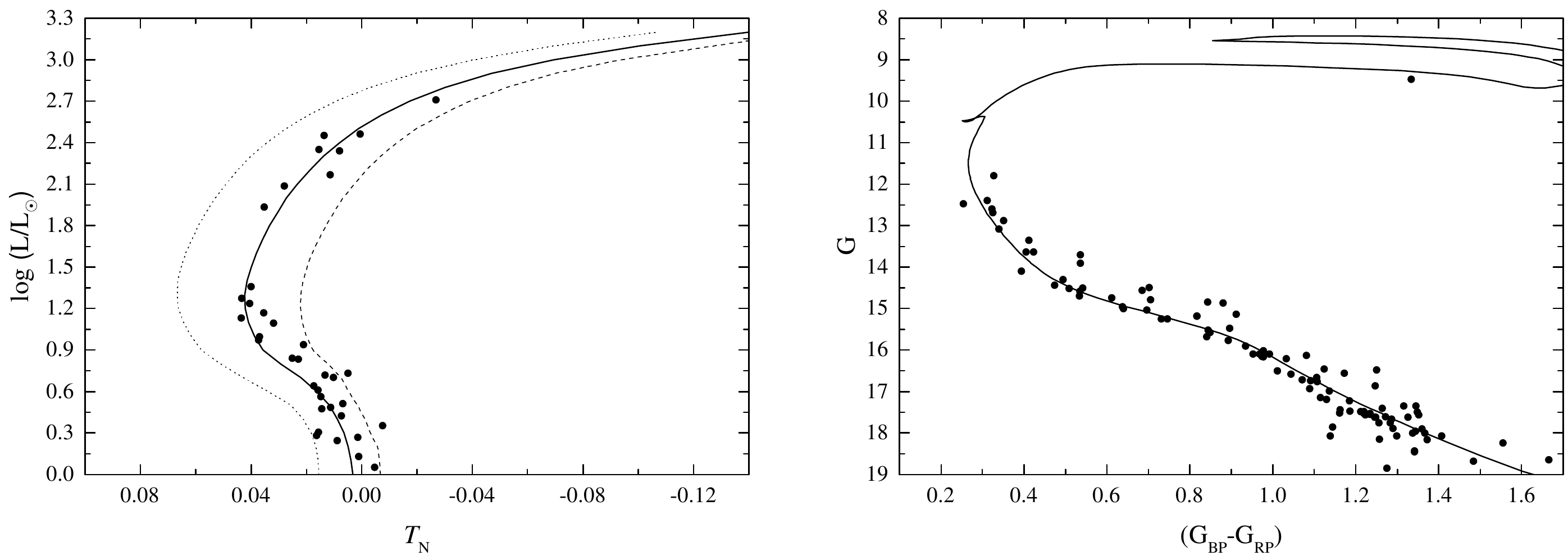}}
\caption{As Figure \ref{fig:basel4}, but for NGC~2186.}
\end{figure*}
% ---------------------------

% ---------------------------
\begin{figure*}
\centering
\resizebox{\hsize}{!}{\includegraphics{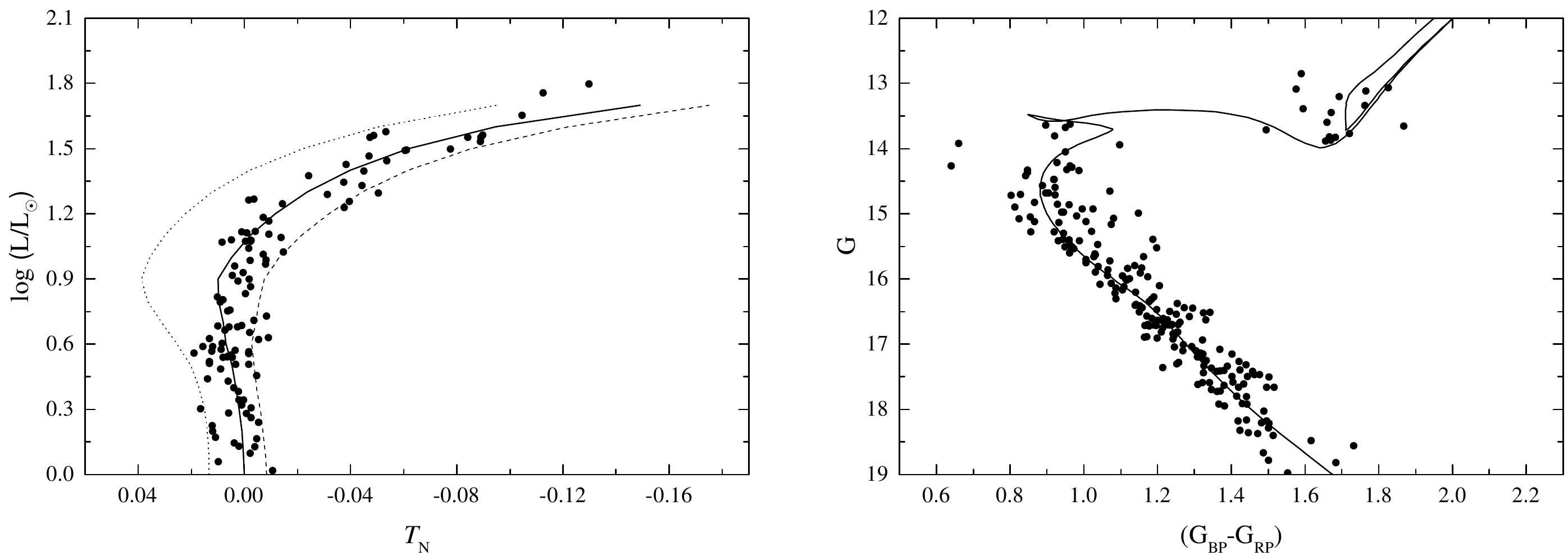}}
\caption{As Figure \ref{fig:basel4}, but for NGC~2236.}
\end{figure*}
% ---------------------------

% ---------------------------
\begin{figure*}
\centering
\resizebox{\hsize}{!}{\includegraphics{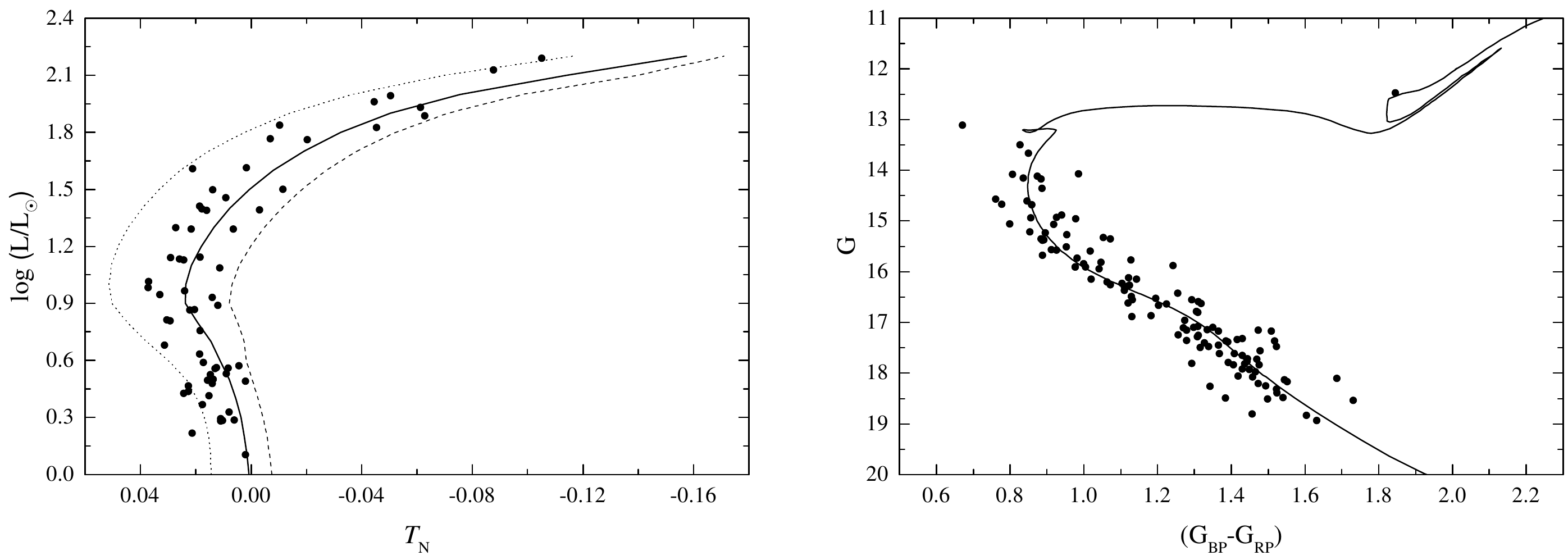}}
\caption{As Figure \ref{fig:basel4}, but for NGC~2259.}
\end{figure*}
% ---------------------------

% ---------------------------
\begin{figure*}
\centering
\resizebox{\hsize}{!}{\includegraphics{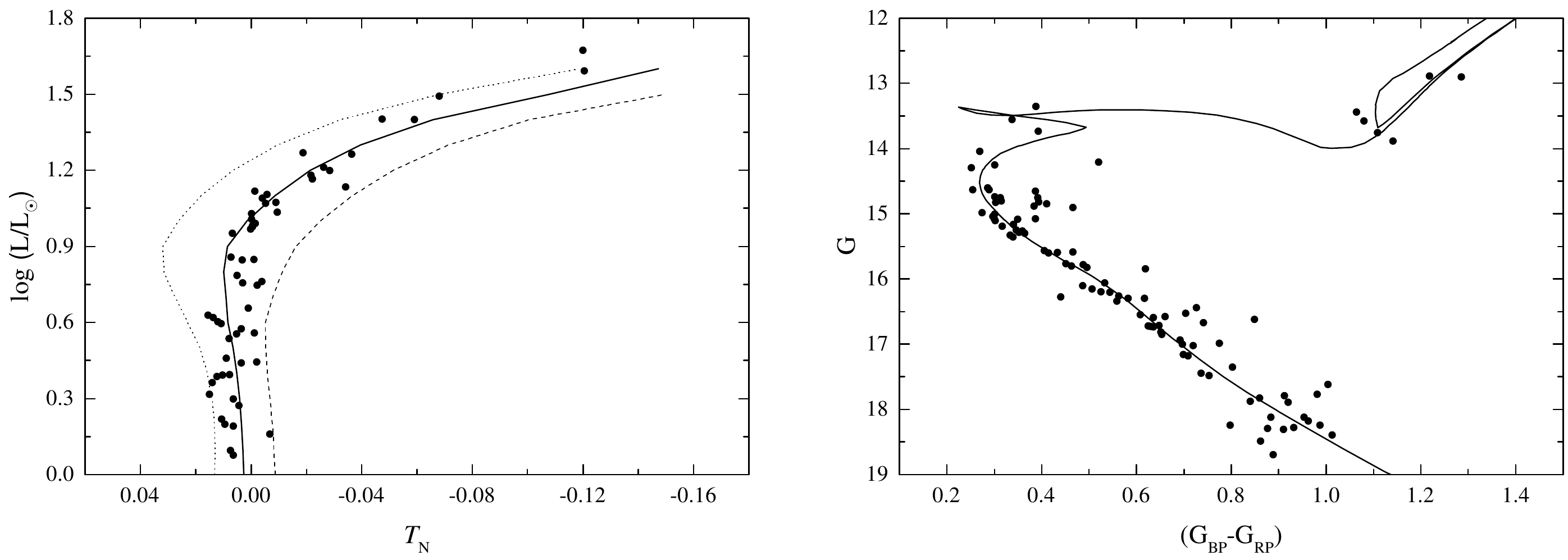}}
\caption{As Figure \ref{fig:basel4}, but for NGC~2304.}
\end{figure*}
% ---------------------------

% ---------------------------
\begin{figure*}
\centering
\resizebox{\hsize}{!}{\includegraphics{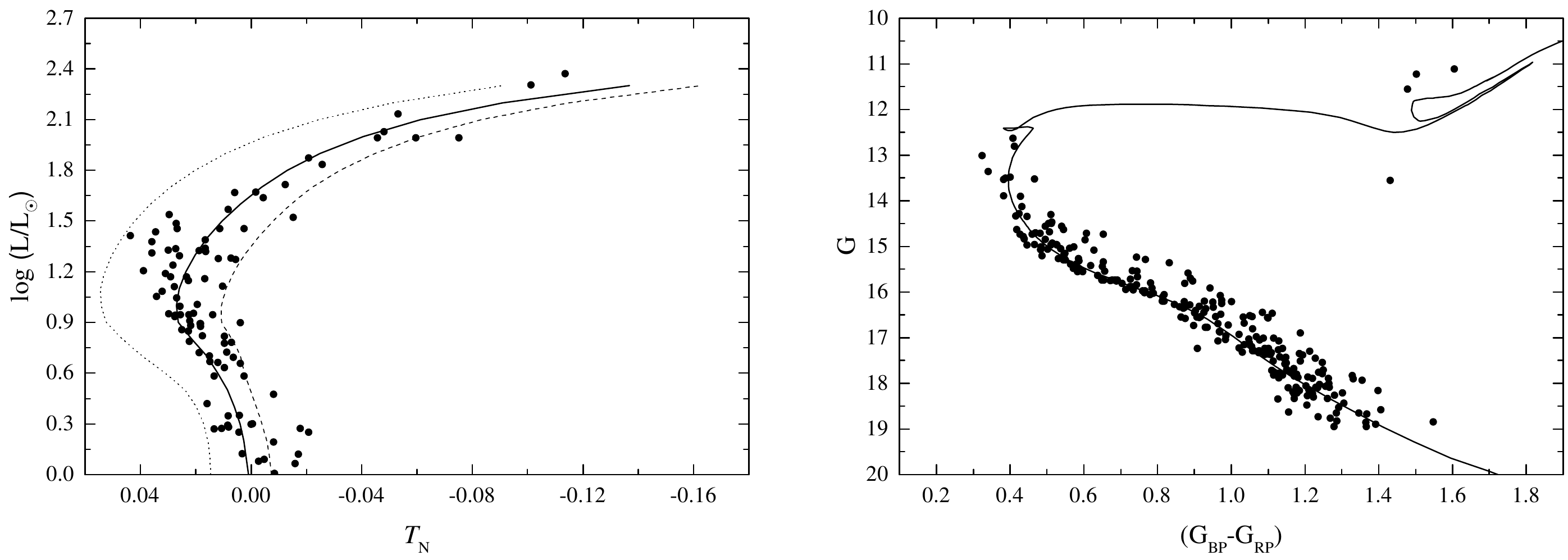}}
\caption{As Figure \ref{fig:basel4}, but for NGC~2383.}
\label{fig:n2383}
\end{figure*}
% ---------------------------

%%%%%%%%%%%%%%%%%%%%%%%%%%%%%%%%%%%%%%%%%%%%%%%%%%

% Don't change these lines
\bsp	% typesetting comment
\label{lastpage}
\end{document}